\documentclass{nature_modifbyHA}

\usepackage{amsmath}
\usepackage{graphicx}
\usepackage{bm}

\title{Spin-Orbit induced phase-shift in Bi$_{2}$Se$_{3}$ Josephson junctions}
\author
{Alexandre Assouline$^{1,4}$, Cheryl Feuillet-Palma$^{1}$, Nicolas Bergeal$^{1}$, Tianzhen Zhang$^{1}$, \\ Alireza Mottaghizadeh$^{1,5}$, Alexandre Zimmers$^{1}$, Emmanuel Lhuillier$^{2}$,\\ Mahmoud Eddrief$^{2}$, Paola Atkinson$^{2}$, Marco Aprili$^{3}$, Herv\'e Aubin$^{1,6}$\\
}

\begin{document} 

\maketitle
\begin{affiliations}
\item{LPEM, ESPCI Paris, PSL Research University; CNRS; Sorbonne Universit\'es, UPMC University of Paris 6,10 rue Vauquelin, F-75005 Paris, France}
\item{Sorbonne Universit\'es, UPMC Univ. Paris 06, CNRS-UMR 7588, Institut des NanoSciences de Paris, 4 place Jussieu, 75005 Paris, France}
\item{Laboratoire de Physique des Solides, CNRS, Univ. Paris-Sud, University Paris-Saclay, 91405 Orsay Cedex, France}

\item{Service de Physique de l'\'Etat Condens\'e, CNRS UMR 3680, IRAMIS, CEA-Saclay, 91191 Gif-sur-Yvette, France}
\item{Current adress:Khatam University, 30 Hakim Azam Street, 1991633356, Tehran, Iran}
\item{Centre de Nanosciences et de Nanotechnologies, CNRS, Univ. Paris-Sud, Universit\'es Paris-Saclay, C2N, 91120 Palaiseau, France}
\end{affiliations}

\begin{abstract}
The transmission of Cooper pairs between two weakly coupled superconductors produces a superfluid current and a phase difference; the celebrated Josephson effect. Because of time-reversal and parity symmetries, there is no Josephson current without a phase difference between two superconductors. Reciprocally, when those two symmetries are broken, an anomalous supercurrent can exist in the absence of phase bias or, equivalently, an anomalous phase shift $\varphi_0$ can exist in the absence of a superfluid current. We report on the observation of an anomalous phase shift $\varphi_0$ in hybrid Josephson junctions fabricated with the topological insulator Bi$_2$Se$_3$ submitted to an in-plane magnetic field. This anomalous phase shift $\varphi_0$ is observed directly through measurements of the current-phase relationship in a Josephson interferometer. This result provides a direct measurement of the spin-orbit coupling strength and open new possibilities for phase-controlled Josephson devices made from materials with strong spin-orbit coupling.
\end{abstract}

\section*{Introduction}
In Josephson junctions\cite{Josephson1962-ey}, the Current-Phase relationship (CPR) is given by the first Josephson equation\cite{Golubov2004-yl}, $I_{\rm J}(\varphi)=I_0\sin(\varphi+\varphi_0)$. Time-reversal and spatial parity symmetries\cite{Rasmussen2016-ct}, P$_x$,P$_y$,P$_z$ impose the equality $I_{\rm J}(\varphi\rightarrow0)=0$ and so only two states for the phase shift $\varphi_0$ are possible. $\varphi_0=0$ in standard junctions and $\varphi_0=\pi$ in presence of a large Zeeman field, as obtained in hybrid superconducting-ferromagnetic junctions\cite{Kontos2001-sl,Guichard2003-cc,Buzdin2005-ja} or in large g-factor materials under magnetic field\cite{Hart2016-vg,Li2017-fv}.

To observe an anomalous phase shift $\varphi_0$ intermediate between 0 and $\pi$, both time-reversal and parity symmetries must be broken\cite{Bezuglyi2002-pz,Braude2007-xn,Buzdin2008-hg,Reynoso2008-qq,Zazunov2009-tt,Grein2009-ro,Malshukov2010-jd,Liu2010-wx,Yokoyama2014-ke,Campagnano2015-lu,Dolcini2015-gv,Bergeret2015-so,Pershoguba2015-hy,Szombati2016-rr}. This can be obtained in systems with both a Zeeman field and a Rashba spin-orbit coupling term H$_{\rm R}= \frac{\alpha}{\hbar} (\mathbf{p}\times \mathbf{e}_z).\bm{\sigma}$ in the Hamiltonian\cite{Buzdin2008-hg,Bergeret2015-so}, where $\alpha$ is the Rashba coefficient, $\mathbf{e}_z$ the direction of the Rashba electric field and $\bm{\sigma}$ a vector of Pauli matrices describing the spin. Physically, these terms lead to a spin-induced dephasing of the superconducting wave-function.

This anomalous phase shift is related to the inverse Edelstein effect observed in metals or semiconductors with strong spin-orbit coupling. While the Edelstein effect consists in the generation of a spin polarization in response to an electric field\cite{Edelstein1990-yj}, the inverse Edelstein effect\cite{Shen2014-lz}, also called spin-galvanic effect, consists in the generation of a charge current by an out-of-equilibrium spin polarization. These two magneto-electric effects are predicted also to occur in superconductors as a consequence of a Lifshitz type term in the free energy\cite{Yip2002-ua,Konschelle2015-ip}. Thus, in a superconductor with a strong Rashba coupling, a Zeeman field induces an additional term in the supercurrent. In Josephson junctions this term leads to the anomalous phase shift\cite{Bergeret2015-so}.

Several designs of Josephson junctions leading to an anomalous phase shift have been proposed theoretically where the Zeeman field can be obtained from an applied magnetic field\cite{Grein2009-ro,Liu2010-wx,Bergeret2015-so} or by using a magnetic element\cite{Buzdin2008-hg}.  These designs include the use of atomic contacts \cite{Reynoso2008-qq}, quantum dots \cite{DellAnna2007-gc,Zazunov2009-tt}, nanowires \cite{Yokoyama2014-ke,Campagnano2015-lu}, topological insulator \cite{Dolcini2015-gv,Alidoust2017-mx,Zyuzin2016-lr,Bobkova2016-oh}, diffusive junction\cite{Bergeret2015-so}, magnetic impurity \cite{Pershoguba2015-hy}, ferromagnetic barrier \cite{Buzdin2008-hg} and diffusive superconducting-ferromagnetic junctions with non-coplanar magnetic texture\cite{Silaev2017-bm}.

Experimentally, the anomalous phase shift  $\varphi_0$ can be detected in a Josephson Interferometer (JI) through measurements of the CPR. Anomalous phase shifts have been identified recently in JIs fabricated from the parallel combination of a normal '0' and '$\pi$' junction\cite{Szombati2016-rr} that breaks the parity symmetries.

Because of its large g-factor $g=19.5$\cite{Wolos2016-lj} and large Rashba coefficient, Bi$_{2}$Se$_{3}$ is a promising candidate for observing the anomalous Josephson effect due to the interplay of the Zeeman field and spin-orbit interaction. In this topological insulator\cite{Zhang2009-ll,Hasan2010-op}, the effective Rashba coefficient of the topological Dirac states is about $\alpha\simeq 3~$eV\AA\cite{Zhang2010-kf}, while the Rashba coefficient of the bulk states, induced by the broken inversion symmetry at the surface, has a value in the range 0.3 - 1.3 eV\AA\, as measured by photoemission\cite{Zhu2011-cx,King2011-sk}.

As detailed in Refs.\cite{Konschelle2015-ip,Bergeret2015-so}, the amplitude of the anomalous phase depends on the amplitude of the Rashba coefficient $\alpha$, the transparency of the interfaces, the spin relaxation terms such as the Dyakonov-Perel coefficient and whether the junction is in the ballistic or diffusive regime. At small $\alpha$, the anomalous phase is predicted proportional to $\alpha^3$, at large $\alpha$, it should be proportional to $\alpha$.

In the ballistic regime\cite{Buzdin2008-hg} and for large $\alpha$, the anomalous phase shift is given by $\varphi_0=\frac{4E_{\rm Z} \alpha L}{(\hbar v_{\rm F})^{2}}$ for a magnetic field of magnitude $B$ and perpendicular to the Rashba electric field, where  $E_{\rm Z}=\frac{1}{2}g\mu_B B$ is the Zeeman energy, $L$ is the distance between the superconductors and $v_{\rm F}$ is the Fermi velocity of the barrier material. For the Rashba spin-split conduction band with $\alpha \approx0.4$~eV\AA, $v_{\rm F}=3.2~10^5$~ms$^{-1}$ and junction length $L=150~$nm, a magnetic field $B=100$~mT generates an anomalous phase $\varphi_0 \simeq 0.01~\pi$, while for Dirac states\cite{Zhang2010-kf} with $v_{\rm F}=4.5~10^5$~ms$^{-1}$, $\varphi_0 \simeq 0.005~\pi$.

In the diffusive regime, the expected anomalous phase shift has been calculated in Ref.\cite{Bergeret2015-so}. For weak $\alpha$, highly transparent interfaces and neglecting spin-relaxation, the anomalous phase shift is given by the relation:

\begin{equation}
    \varphi_0=\frac{\tau m^{*2} E_{\rm Z}(\alpha L)^3}{3\hbar^6 D}
\end{equation}

where $\tau=0.13~$ps is the elastic scattering time, $D=\frac{1}{3}v_{\rm F}^2\tau=40~$cm$^2$s$^{-1}$ is the diffusion constant and $m^*=0.25~m_e$ is the effective electron mass\cite{Hyde1973-oo}.

To test these theoretical predictions, we fabricated single Josephson junctions and JIs from Bi$_{2}$Se$_{3}$ thin films of 20~quintuple layers thick, $\sim 20~$nm, grown by Molecular Beam Epitaxy and protected by a Se layer, see Supplementary Note 1 and Supplementary Figure 1. As described in Supplementary Figure 2, these junctions are in the diffusive regime. 

From the measurement of the relative phase shift between two JIs with different orientations of the Josephson junctions with respect to the in-plane magnetic field, we observed unambiguously the anomalous phase-shift predicted by Eq.~1.

\section*{Results}

\subsection{Current-Phase Relationship and Shapiro steps}

\begin{figure*}
	\begin{center}
		\includegraphics[width=14cm]{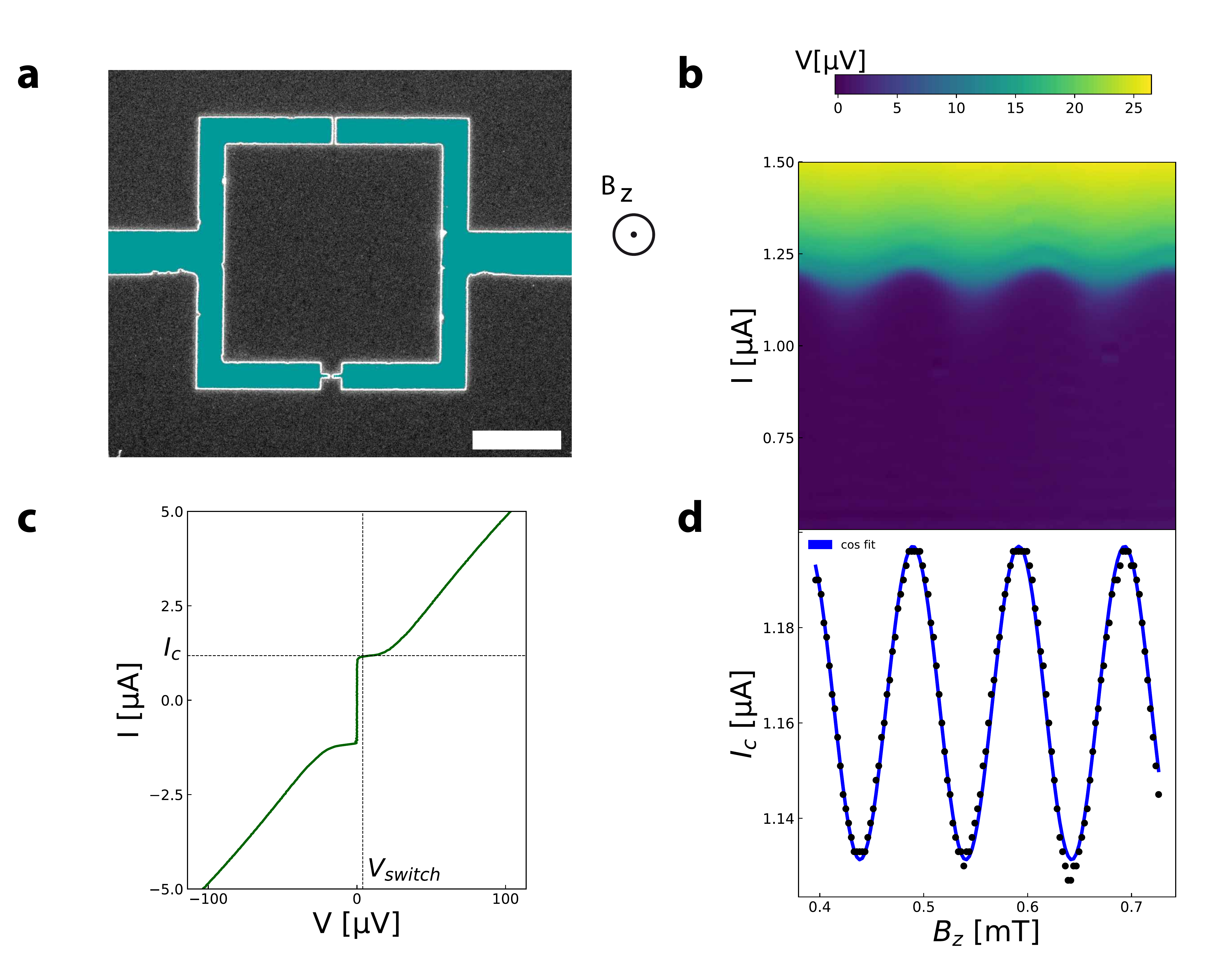}
		\caption{\label{Fig1} \textbf{Probing the CPR with a Josephson interferometer.} \textbf{a} SEM image of a JI device. The scale bar is 2~$\mu$m. This device consists of two Josephson junctions in parallel and enables probing the CPR of the smaller junction. The Aluminum electrodes are 50 nm thick. \textbf{b} Voltage map as a function of current and magnetic field $B_z$ applied perpendicular to the film plane showing zero voltage below a critical current value. The critical current of the small junction oscillates with the magnetic flux in the superconducting loop. \textbf{c} IV curve of the JI device. The critical current value $I_{\rm c}$ is extracted when the junction develops a finite voltage defined as $V_{\rm switch}$ and indicated by the dashed vertical line. \textbf{d} Critical current extracted from (c) as a function of $B_z$. The oscillations are properly described by the function $I_{\rm c}(B_z)=I_{\rm c_1}+I_{\rm c_2}\cos(\omega B_z)$, indicating a sinusoidal CPR as expected for conventional Josephson junctions.}
	\end{center}
\end{figure*}

The JI shown in Fig.~\ref{Fig1}a consists of two junctions in parallel of widths $W_1=600~$nm and $W_2=60~$nm, respectively. The phase differences $\varphi_1$ and $\varphi_2$ for the two junctions are linked by the relation $\varphi_1-\varphi_2=2\pi\frac{\phi}{\phi_0} $, where $\phi=B_zS$ is the magnetic flux enclosed in the JI of surface S, $B_z$ is a small magnetic field perpendicular to the sample, i.e. along $\mathbf{e_z}$, and $\phi_0$ is the flux quantum. In this situation, the Zeeman energy is negligible and oriented along the Rashba electric field, which implies that $\varphi_0=0$. As the critical current $I_{\rm c_1}$ is much higher than $I_{\rm c_2}$, then $\varphi_1=\pi/2$ and $I_{\rm c}=I_{\rm c_1}+I_{\rm c_2}\cos(\omega B_z)$ with $\omega=2\pi S/\phi_0$\cite{Della_Rocca2007-dv}. Thus, a measurement of the critical current $I_{\rm c}$ as function of $B_z$ provides a measure of the current $I_{\rm c_2}$ as function of $\varphi_2$, i.e. the CPR. From the voltage map as a function of current $I$ and $B_{z}$, shown Fig.~\ref{Fig1}b, the critical current $I_{\rm c}$ is extracted when the voltage across the device exceeds the value $V_{\rm switch}=4~\mu$V, as shown in Fig.~\ref{Fig1}c. We find that the CPR displays a conventional sinusoidal form $I_{\rm J}=I_{\rm c}\sin(\varphi)$, as shown by the fit in Fig.~\ref{Fig1}d. Furthermore, under microwave irradiation, the JI displays a conventional, $2\pi$ periodic, Shapiro pattern, as shown  Fig.~\ref{Fig2}, and detailed discussion in Supplementary Note 2.

\subsection{Asymmetric Fraunhofer pattern with in-plane magnetic field}

Fig.~\ref{Fig3}bcd show resistance maps dV/dI of a single junction, Fig.~\ref{Fig3}a, as function of current $I$ and $B_z$ for different values of an in-plane magnetic field, $B_y$. Fig.~\ref{Fig3}e shows the corresponding critical current curves. A Fraunhofer pattern is observed with the first node located at $B_0\simeq1.2$~mT. This value is consistent with the theoretical value $B_0=\frac{\phi_0}{W(L+2\lambda_{z})}$, using the effective magnetic penetration depth $\lambda_{z}=175$~nm and taking into account flux-focusing effects, see Supplementary Note 3. While for $B_y=0$, the Fraunhofer pattern is symmetric with respect to $B_z$, this pattern becomes asymmetric upon increasing the amplitude of $B_y$. This evolution is shown in the critical current map as a function of $B_z$ and $B_y$ in Fig.~\ref{Fig3}f. We observed a much less pronounced asymmetry when we apply the magnetic field in the $x$ direction as shown in Supplementary Figure 5. Similar behavior has been observed recently in InAs\cite{Suominen2017-ls} interpreted as the consequence of a combination of spin-orbit, Zeeman and disorder effects. As described in Ref.[\cite{Rasmussen2016-ct}], the generation of an anomalous phase shift requires breaking all symmetry operations U leaving UH$(\varphi)$U$^\dagger$=H$(-\varphi)$, where H is the full Hamiltonian of the system including spin-orbit interactions. These symmetry operations are shown in Table 1 together with the parameters breaking those symmetries. This table shows that for a system with a finite spin-orbit coefficient $\alpha$, finite $B_y$ is sufficient to generate an anomalous phase shift. However, additional symmetry operations U leaving UH$(B_z,\varphi)$U$^\dagger$=H$(-B_z,\varphi)$ must be broken to generate an asymmetric Fraunhofer pattern, as shown in Supplementary Table 1. In addition to non-zero values for $\alpha$ and $B_y$, disorder along y direction, i.e. non-zero $V_y$, is required to generate an asymmetric Fraunhofer pattern. AFM images, as in shown in Supplementary Figure 6g, show that the MBE films present atomic steps. Because of the dependence of the Rashba coefficient on film thickness\cite{Zhang2010-kf}, phase jumps along the y direction of the junction can be produced by jumps in the Rashba coefficient and explains the polarity asymmetry of the Fraunhofer pattern. As detailed in Supplementary Note 4, using a simple model, the asymmetric Fraunhofer pattern measured experimentally can be simulated, as shown in Fig.~\ref{Fig3}g.

\subsection{Current-Phase Relationship with in-plane magnetic field}

To unambiguously demonstrate that an anomalous phase shift $\varphi_0$ can be generated by finite spin-orbit coefficient $\alpha$ and finite magnetic field $B_y$ alone, a direct measurement of the CPR with in-plane magnetic field is required. To that end, we measured simultaneously two JIs, oriented as sketched in Fig.~\ref{Fig4}a, differing only by the orientation of the small junctions with respect to the in-plane magnetic field. 

The CPRs for the two JIs are measured as function of a magnetic field making a small tilt $\theta$ with the sample plane, which produces an in-plane $B_y=B\cos(\theta)$ and a perpendicular $B_z=B\sin(\theta)$ magnetic field, as sketched in Fig.~\ref{Fig4}c. In this situation, the critical current for the reference JI changes as $I_{\rm c}\propto\cos(\omega_{\rm ref} B)$ with $\omega_{\rm ref}=\frac{2\pi S_{\rm ref}}{\phi_0}\sin(\theta)$ where $S_{\rm ref}$ is the surface of the JI. For the anomalous JI, the critical current changes as $I_{\rm c}\propto\cos(\omega B)$ with :

\begin{equation}
\omega=\frac{2\pi S}{\phi_0}\sin{\theta}+C_{\varphi_0}\cos(\theta)
\label{JIphase}
\end{equation}
where $C_{\varphi_0}=\frac{\tau m^{*2} g\mu_{\rm B} (\alpha L)^3}{6\hbar^6 D}$ in the diffusive regime.

In Eq.\ref{JIphase}, the first term arises from the flux within the JI of area S, the second term arises from the anomalous phase shift $\varphi_0=C_{\varphi_0}B\cos(\theta)$.

Fig.~\ref{Fig4}b shows voltage maps for two different orientations $\theta$, Fig.~\ref{Fig4}c. At low $B$, the two JIs are in-phase and become out-of-phase at higher magnetic field, indicating that the frequency $\omega$ of the anomalous JI is slightly larger than the reference JI. This is also visible on the critical current plot, Fig.~\ref{Fig5}a, extracted from these voltage maps. To see this more clearly, the average critical current, shown as a continuous line in Fig.~\ref{Fig5}a, is removed from the critical current curve and the result shown in Fig.~\ref{Fig5}b for the two JIs. On these curves, the nodes at $\pi(2n+1/2)$, $n=0,1,..$, are indicated by large red (blue) dots for the reference (anomalous) curve. At low magnetic field, the two JIs are in-phase as indicated by the blue and red dots being located at the same field position. Upon increasing the in-plane magnetic field, the two JIs become out-of-phase with the anomalous JI oscillating at a higher frequency than the reference JI, as indicated by the blue dot shifting to lower magnetic field position with respect to the red dot. This increased frequency for the anomalous device is expected from Eq.~\ref{JIphase} as a consequence of the anomalous phase shift. Supplementary Figure 7ab shows additional data taken from negative to positive magnetic field, across zero magnetic field. A plot of the phase difference between the two JIs as function of in-plane magnetic field, shown in Supplementary Figure 7c, demonstrates that the two JIs are in-phase at zero magnetic field and reach a dephasing approaching about $\pi/2$ for an in-plane magnetic field of $\simeq 80$~mT.

One also sees that the oscillation period of both JIs increase with increasing $B$. As detailed in Supplementary Note 3, this is due to flux focusing that makes the effective area of the JIs larger at low magnetic field. As the effect of flux focusing decreases with the increasing penetration depth at higher magnetic field, the effective areas of the JIs decreases upon increasing the magnitude of the magnetic field and so the period of oscillations increases.

While the two JIs have been fabricated with nominally identical areas, to exclude that the observed difference in frequencies between the two JIs is due to a difference of areas, we plot in Fig.~\ref{Fig5}c, the frequency ratio $\frac{\omega}{\omega_{\rm ref}}(\theta)$ measured at different angles $\theta$. Because each curve contains several periods $T_{i}$, the frequency ratio is obtained from the average between the $N$ periods ratio as $\omega/\omega_{\rm ref}=\frac{1}{N}\sum_{i=1}^{N}\frac{T_{i,\rm ref}}{T_i}$, where $T_{i,\rm ref}$ and $T_i$ are the i$^{th}$ oscillation period for the reference and for the anomalous device respectively. This method enables ignoring the flux focusing effect because the ratio is only taken between two periods measured at about the same magnetic field. We find that the experimental data follows the relation:

\begin{equation}
\frac{\omega}{\omega_{\rm ref}}(\theta)=\frac{S}{S_{\rm ref}}+\frac{C_{\varphi_0}\phi_0}{2\pi S_{\rm ref}\tan(\theta)}
\label{ratiofrequency}
\end{equation}

At large $\theta$, this ratio is equal to the ratio of areas $S/S_{\rm ref}\simeq 1$, however, for small $\theta$, this ratio increases as $1/\tan(\theta)$, indicating the presence of an anomalous phase shift $\varphi_0$.

Another way of extracting the frequency is described in the Supplementary Note 5 and leads to the same result, as shown in Supplementary Figure 8.

\section*{Discussion}

A fit of the experimental data with Eq.~\ref{ratiofrequency}, Fig.~\ref{Fig5}c, enables extracting the coefficient $\frac{C_{\varphi_0}\phi_0}{2\pi S_{\rm ref}}=41\pm5~10^{-5}$. Using the expression of $C_{\varphi_0}$ in the diffusive regime given above, we calculate a spin-orbit coefficient $\alpha=0.38\pm0.015$~eV\AA. This value of the Rashba coefficient is consistent with the value extracted from Rashba-split conduction band observed by photoemission measurements\cite{Zhu2011-cx,King2011-sk}. Table 2 gives the anomalous phase shift extracted from the critical current oscillations at the largest magnetic field about 80-100~mT. The phase shift is extracted from the magnetic field difference between the last nodes of the oscillations, indicated by blue and red dots on Fig.~5. At this largest magnetic field, we find an anomalous phase shift $\varphi_0\simeq0.9\pi$ for all three tilt angles $\theta$. This shows that the anomalous phase shift depends only on the parallel component of the magnetic field as expected. This experimental value is compared with the theoretical values calculated in the ballistic regime, for the Rashba-split conduction states and Dirac states, and in the diffusive regime, for the Rashba-split conduction states. This table shows that the Dirac states provide only a phase shift of $0.005\pi$ and so cannot explain the experimental data. The table also shows that the Rashba-split conduction states provide a phase shift of only $0.01\pi$ in the ballistic regime while they provide an anomalous shift of $0.94 \pi$ in the diffusive regime, close to the experimental value, confirming that the junctions are indeed in the diffusive regime and demonstrating the validity of the theory leading to Eq.~1.

A detailed look at Table 1 shows that the anomalous shift observed here must be the consequence of finite Rashba coefficient and in-plane magnetic field. While Table 1 shows that disorder alone $V_y$ is sufficient to generate an anomalous phase shift, this disorder-induced anomalous phase shift should exist even at zero magnetic field and should not change with magnetic field. In contrast, as discussed above, we have seen that the two JIs are in-phase near zero magnetic field and become out of phase only for at finite magnetic field. Thus, this observation implies that disorder $V_y$ is absent, which is plausible as the small Josephson junction is only 150~nm$\times$150~nm. In the absence of disorder $V_y$, Table 1 shows that the only way for an anomalous phase shift to be present is that the coefficient $\alpha$ be non-zero. Indeed, if $\alpha$ were zero, the first and third symmetry operations of Table 1 would not be broken even with finite $B_y$.

To summarize the result of this work, the simultaneous measurements of the CPR in two JIs making an angle of 90$^{\circ}$ with respect to the in-plane magnetic field enabled the identification of the anomalous phase shift $\varphi_0$ induced by the combination of the strong spin-orbit coupling in Bi$_{2}$Se$_{3}$ and Zeeman field. This anomalous phase shift can be employed to fabricate a phase battery, a quantum device of intense interest for the design and fabrication of superconducting quantum circuits\cite{Ortlepp2006-tb,Feofanov2010-ca}.

\begin{methods}
\subsection{Sample preparation}
The Bi$_{2}$Se$_{3}$ samples were grown by Molecular Beam Epitaxy. The crystalline quality of the films was monitored in-situ by reflection high energy electron diffraction and ex situ by x-ray diffraction, and by post growth verification of the electronic structure though the observation of the Dirac cone fingerprint in angle-resolved photoemission spectra as described in Ref.~\cite{Vidal2013-tc}. Following growth, the samples were capped with a Se protective layer. The Josephson junctions are fabricated on these thin films with standard e-beam lithography, e-beam deposition of Ti(5nm)/Al(20-50~nm) electrodes and lift-off. The Se capping layer is removed just before metal evaporation by dipping the samples in a NMF solution of Na$_{2}$S. In the evaporator chamber, the surface is cleaned by moderate in-situ ion beam cleaning of the film surface before metal deposition. While for standard junctions, an aluminum layer 50~nm thick is usually deposited, we also fabricated junctions with 20~nm thick electrodes to increase their upper critical field as required by the experiments with in-plane magnetic field. See Supplementary Figure 3 for a lateral sketch of the devices. After microfabrication, the carrier concentration is about $10^{19}~$cm$^{-3}$ and the resistivity about 0.61~m$\Omega$.cm, as shown in Supplementary Figure 2. A comparison between the normal state junction resistance of the order of 20-50~$\Omega$ and the resistivity of the films indicates negligible contact resistance, i.e. the junction resistance is due to the Bi$_{2}$Se$_{3}$ film between the electrodes.

\subsection{Measurements details}
The values for the normal resistance and critical current values measured on 20 devices are found to be highly reproducible, demonstrating the reliability of our procedure for surface protection and preparation before evaporation of the electrodes. The devices are measured in a dilution fridge with a base temperature of 25~mK. The IV curves are measured with standard current source and low noise instrumentation amplifiers for detecting the voltage across the junctions. The measurement lines are heavily filtered with $\pi$ filters at room temperature at the input connections of the cryostat. They are also filtered on the sample stage at low temperature with 1 nF capacitances connecting the measurements lines to the ground.
\end{methods}

\section*{Data availability}
The data that support the main findings of this study are available from the corresponding author upon request. The source data underlying Fig.~5c and Supplementary Fig.~8d are provided as Source Data files.

\section*{References}

\begin{addendum}
	\item[Acknowledgements] The devices have been made within the consortium Salle Blanche Paris Centre. We acknowledge fruitful discussions with S. Bergeret and J. Danon. This work was supported by French state funds managed by the ANR within the Investissements d'Avenir programme under reference ANR-11-IDEX-0004-02, and more specifically within the framework of the Cluster of Excellence MATISSE. We also thanks L. Largeau (C2N: Centre de Nanosciences et de Nanotechnologies-Université Paris-Sud) and D. Demaille (INSP : Institut des NanoSciences de Paris-Sorbonne Univerité)) for the atomic-resolved HAADF-STEM images.

	\item[Author contributions] H.A. proposed and supervised the project. M.E. and P.A. have grown the Bi$_2$Se$_3$ thin films by MBE and did the structural characterization (AFM, X-Ray, TEM). A.A. designed and microfabricated the samples with the help of C.F.P., T.Z., A.M., A.Z., E.L., M.A. and H.A.. A.A, M.A. and H.A made the measurements with the help of C.F.P., N.B.. A.A., M.A. and H.A. analyzed the data and wrote the manuscript.
	
	\item[Competing Interests] The Authors declare no competing interests.
	
	\item[Correspondence] Requests for materials should be addressed to H.A. (herve.aubin@c2n.upsaclay.fr) and A.A. (alexandre.assouline@cea.fr).

\end{addendum}

\section*{Figure legends}
\clearpage

\clearpage

\begin{figure*}
	\begin{center}
		\includegraphics[width=14cm]{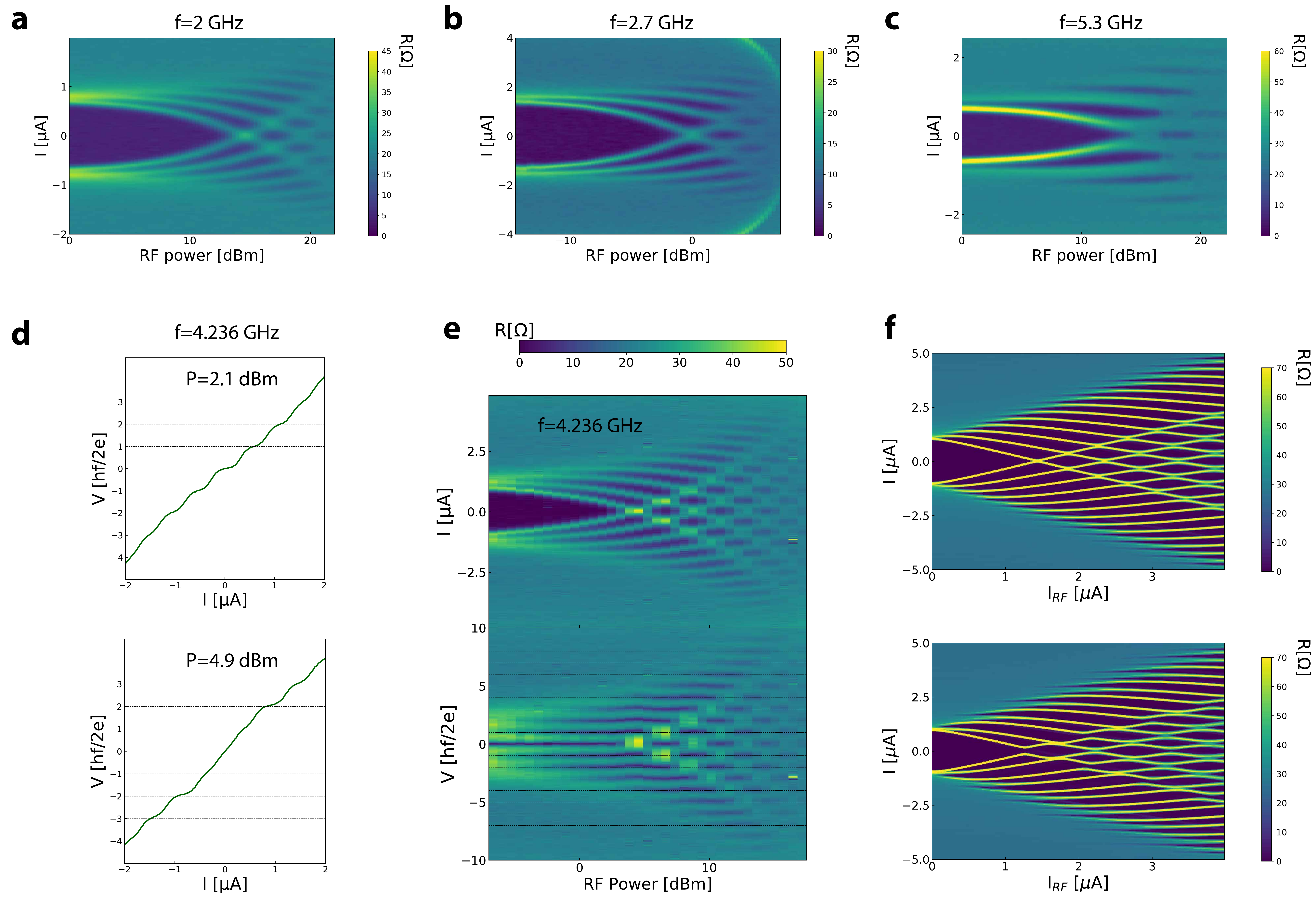}
		\caption{\label{Fig2} \textbf{A.C. Josephson effect in Bi$_2$Se$_3$.} For this Josephson junction, 50 nm thick Aluminum electrodes are used. \textbf{abc} Resistance maps as a function of current and RF power for different microwave frequencies (a) f=2~GHz, (b) 2.7~GHz and (c) 5.3~GHz. The zero resistance regions correspond to voltage plateaus. \textbf{d} IV curves showing the Shapiro steps for two values of the microwave power. The n$^{th}$ current step appears at a voltage $V_n=\frac{nhf}{2e}$. \textbf{e} Resistance map as a function of current (upper panel) or voltage (lower panel) and RF power at the frequency f=4.236~GHz. In the lower panel as a function of voltage, dashed lines are plotted at $V_n=\frac{nhf}{2e}$. \textbf{f} Theoretical maps calculated at f=4.236~GHz with the RSJ model, see Supplementary Note 2, using two different CPRs: A conventional 2$\pi$ periodic CPR $I_{\rm J}=I_{\rm c}\sin(\varphi)$ is used for upper panel, it reproduces properly the experimental data. An unconventional CPR $I_{\rm J}=I_{\rm c}(\frac{4}{5}\sin(\varphi)+\frac{1}{5}\sin(\varphi/2))$ is used for the lower panel where the odd steps have a lower current amplitude.}
	\end{center}
\end{figure*}

\begin{figure*}
	\begin{center}
		\includegraphics[width=14cm]{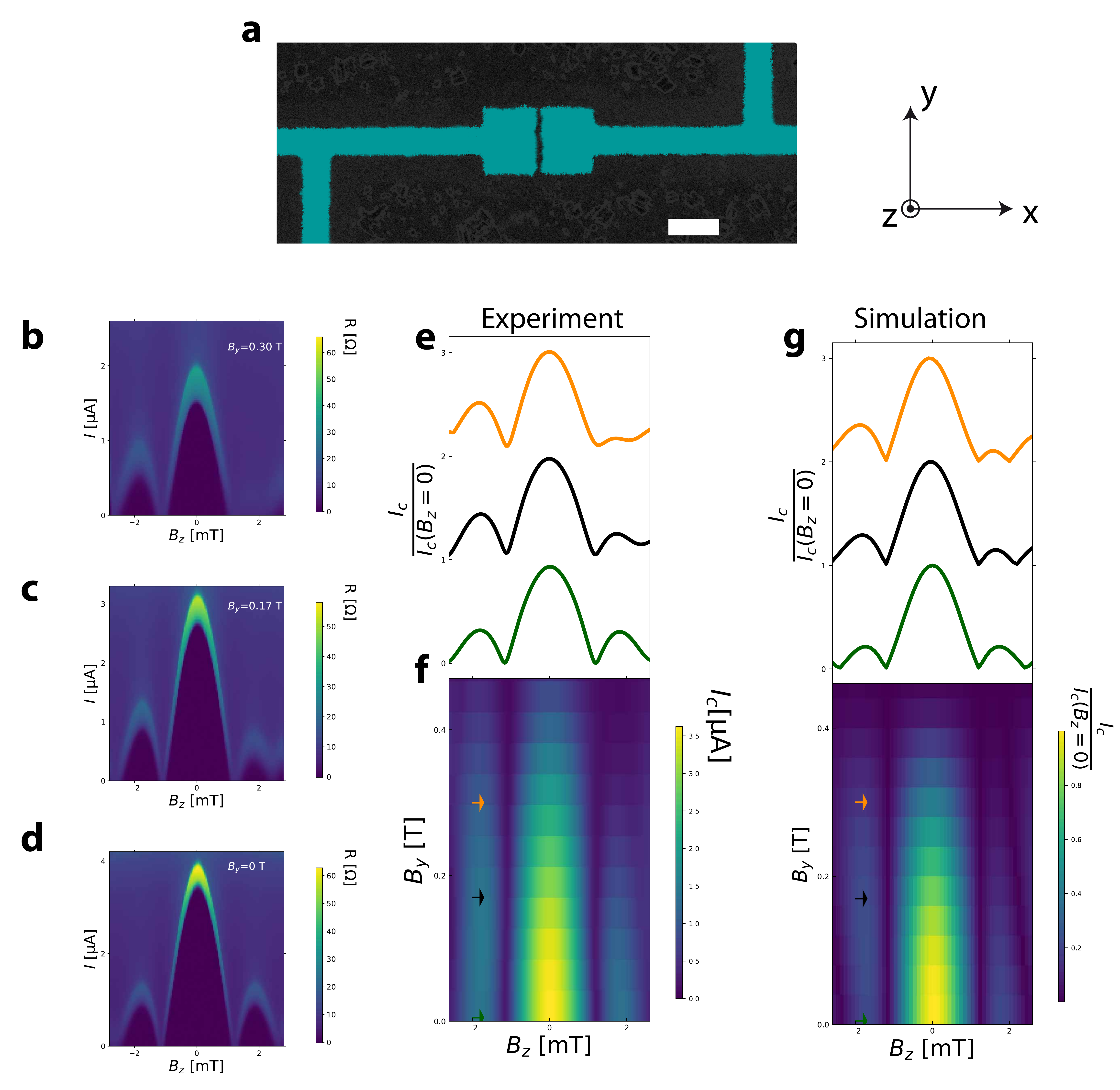}
		\caption{\label{Fig3} \textbf{Asymmetric Fraunhofer pattern.} \textbf{a} SEM image of a single junction device with 20 nm thick Aluminum electrodes. The scale bar is 2~$\mu$m. The spacing between the two superconducting electrodes is L=150 nm and the width of the junction W=2 $\mu$m. \textbf{bcd} Resistance maps as function of current and perpendicular magnetic field $B_z$, for different values of the in-plane magnetic field $B_y$. The corresponding critical current curves are shown panel e. At zero in-plane magnetic field, panel d, the Fraunhofer pattern is symmetric with respect to $B_z$. The pattern becomes asymmetric upon increasing $B_y$. \textbf{f} Critical current map as a function of $B_z$ and $B_y$ showing the evolution from symmetric Fraunhofer pattern at $B_y=0$ to the asymmetric Fraunhofer pattern for $B_y>0$. \textbf{g} Numerical simulation of the anomalous Fraunhofer pattern due to disorder and spin-orbit coupling. The model is described in the Supplementary Note 4.}
	\end{center}
\end{figure*}

\begin{figure*}
	\begin{center}
		\includegraphics[width=14cm]{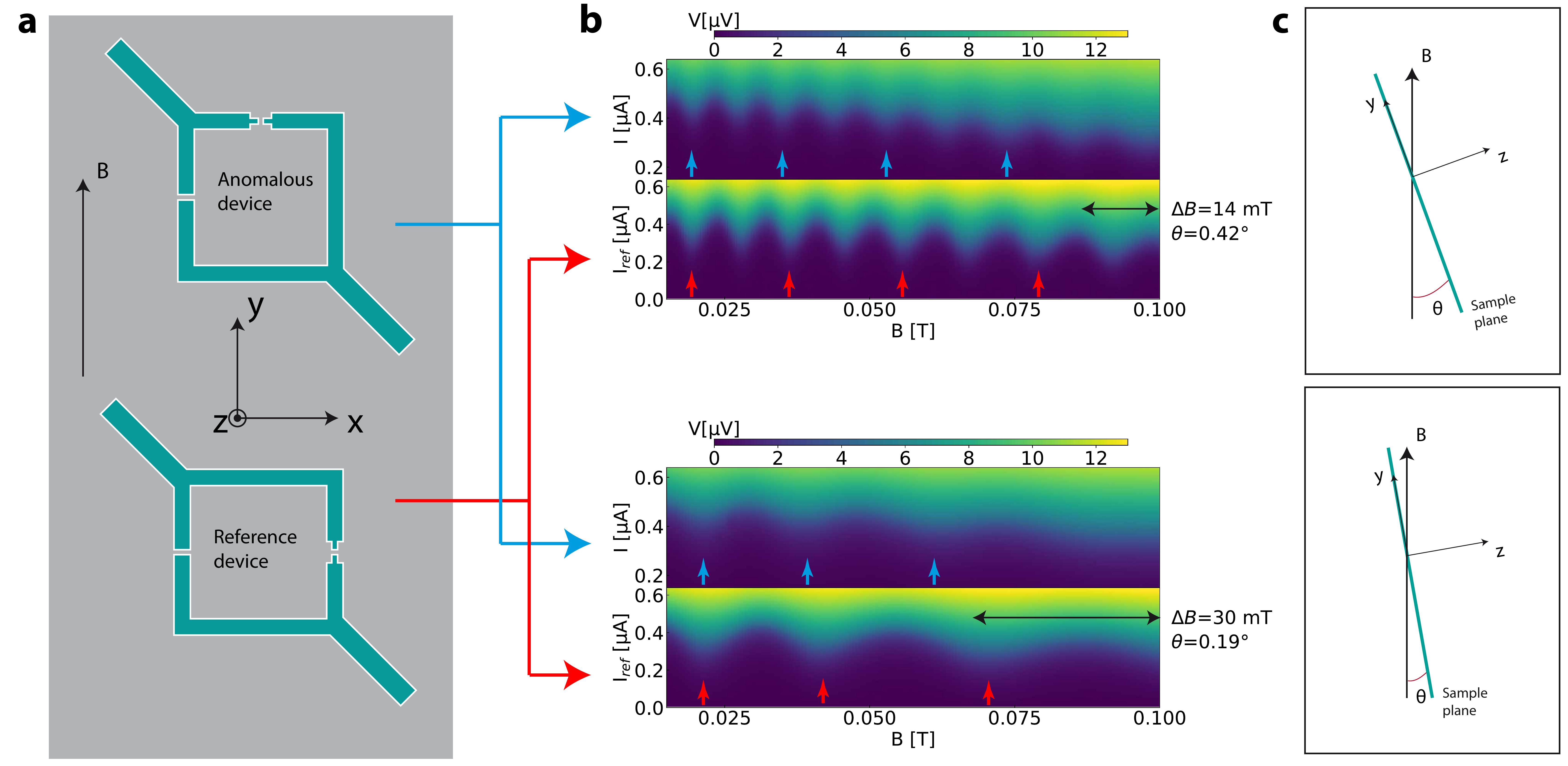}
		\caption{\label{Fig4} \textbf{Probing the anomalous phase shift.} \textbf{a} Sketch of the setup consisting of two JIs fabricated on the same chip. The reference and anomalous JIs have identical area, respectively $S$ and $S_{\rm ref}$, where $S\simeq S_{\rm ref}\simeq20.6~\mu$m$^2$. The Aluminum electrodes are 20 nm thick. From table 1, in the absence of disorder, an anomalous phase shift induced by Rashba spin-orbit coupling can be generated only by an in-plane magnetic field B$_y$. \textbf{b} Voltage map showing the critical current oscillation of the two devices as a function of magnetic field B. The critical current of both devices oscillate due to the perpendicular component of the magnetic field $B_z=B \sin(\theta)$, as sketched panel \textbf{c}. The oscillation frequency can be changed by mechanically tilting the sample, i.e by changing the angle $\theta$ between the plane containing the superconducting loop and the magnetic field B. The frequency of the anomalous device is larger than the reference as a consequence of the anomalous phase shift. The colored arrows are guide to the eyes, to help visualizing the increased phase shift in the anomalous device.
		}
	\end{center}
\end{figure*}

\begin{figure*}
	\begin{center}
		\includegraphics[width=8cm]{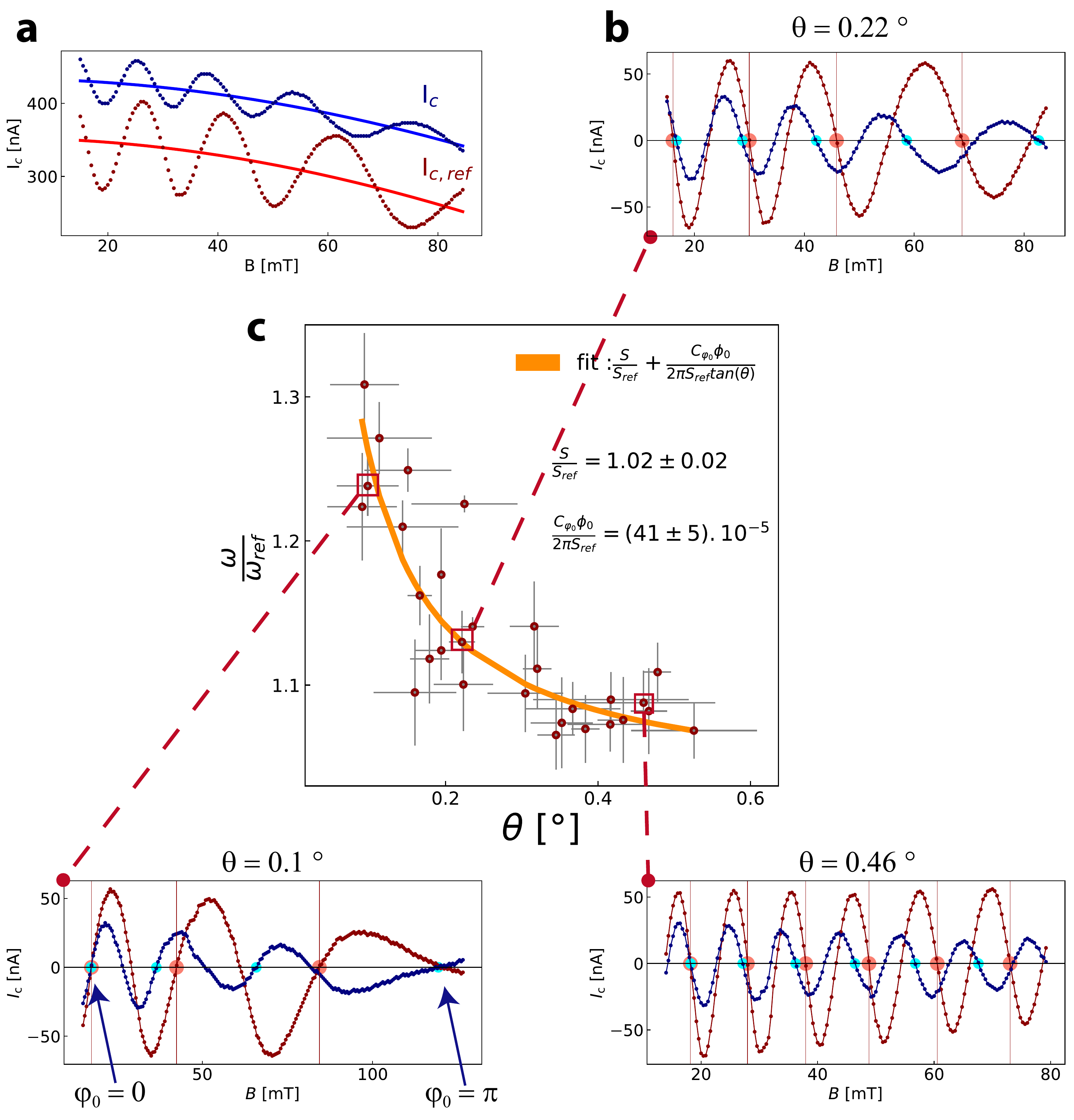}
		\caption{\label{Fig5} \textbf{Comparison of the JIs frequencies as a function of the angle $\theta$.} \textbf{a} The critical current is extracted and shown as a function of magnetic field. The red and blue curves correspond to the reference and anomalous device, respectively. As the critical current of the large junction decreases with the magnetic field, this leads to a decreasing background fitted for both devices and shown by the continuous lines. \textbf{b} Critical current as a function of magnetic field where the background, indicated by continuous lines in panel a, are subtracted. The anomalous device shows a larger oscillation frequency than the reference device. The colored dots help to visualize the increased frequency of the anomalous device (blue dots) with respect to the reference device (red dots). The two JIs are in-phase at low magnetic field and become out-of-phase at high magnetic field, with an anomalous phase shift reaching $\varphi_0\simeq\pi$ for $B\simeq 100$~mT. The periods are compared one by one and the average of the single period ratios provides the value $\frac{\omega}{\omega_{\rm ref}}$. \textbf{c} The ratio of the oscillation frequencies is plotted as a function of the angle $\theta$. Without the generation of an anomalous phase, this ratio should be constant and equal to the surface ratio $\frac{S}{S_{\rm ref}}\simeq1$. This ratio diverges as $\frac{1}{\theta}$ for small $\theta$, as shown by Eq.~\ref{ratiofrequency}. Fitting the curve with Eq.~\ref{ratiofrequency} provides the spin-orbit coefficient $\alpha$.}
	\end{center}
\end{figure*}

\clearpage

\section*{Tables}
\clearpage

\begin{table}
\begin{center}

\begin{tabular}{|l|c|}
  \hline
  UH$(\varphi)$U$^\dagger$=H$(-\varphi)$ & Broken by \\
  \hline
  $P_yP_x$ & $\alpha$, $V_x$, $V_y$  \\
  $\sigma_zP_yP_x$ & B$_x$, B$_y$, $V_x$, $V_y$  \\
  $\sigma_xP_yT$ & B$_x$, $\alpha$, $V_y$  \\
  $\sigma_yP_yT$ & B$_y$, $V_y$  \\

  \hline
\end{tabular}

\caption{\textbf{Symmetry operations U protecting $H(\varphi)=H(-\varphi)$ from Ref.\cite{Rasmussen2016-ct}}. The symmetry operations on the left column are broken by one of the parameters on the right column. These parameters include the in-plane magnetic fields $B_x,B_y$, the asymmetric disorder potentials $V_x,V_y$ and the spin-orbit term $\alpha$ which is the consequence of the structural inversion asymmetry (Rashba). To generate an anomalous phase $\varphi_0$, each symmetry operator, one per line of the table, must be broken. For example, the combination of the magnetic field $B_y$ and the spin-orbit coupling $\alpha$ is enough to break all symmetries.}
\label{table_paramI}
\end{center}
\end{table}

\begin{table}
\begin{center}

\begin{tabular}{|l|c|c|c|c|c|c|}
  \hline
   & $\theta=0.1^{\circ}$& $\theta=0.22^{\circ}$ &$\theta=0.46^{\circ}$ & Ballistic & Dirac & Diffusif \\
  \hline
  $\varphi_0$ &0.88$\pi$ & 1.01$\pi$ & 0.85$\pi$& 0.01$\pi$&0.005$\pi$& 0.94$\pi$\\
  \hline
\end{tabular}

\caption{\textbf{Anomalous phase shifts obtained at $B\simeq$100~mT, compared to theory} The three first columns show the anomalous phase shifts extracted from the last nodes of the critical current oscillations shown in Fig.~5 for the three curves taken at different angles $\theta$. The last three columns show the calculated anomalous phase shifts in the ballistic regime, for the Rashba-split conduction states and Dirac states, and in the diffusive regime, for the Rashba-split conduction states. These theoretical values are calculated using $\alpha=0.4$~eV\AA~for the Rashba-split conduction states and $\alpha=3$~eV\AA~for the Dirac states. See main text for the formula and other parameters.}
\label{table_dephasage}
\end{center}
\end{table}

\end{document}


\maketitle

\clearpage

\section*{Supplementary Figures}

\begin{figure*}[ht!]
	\begin{center}
		\includegraphics[width=16cm]{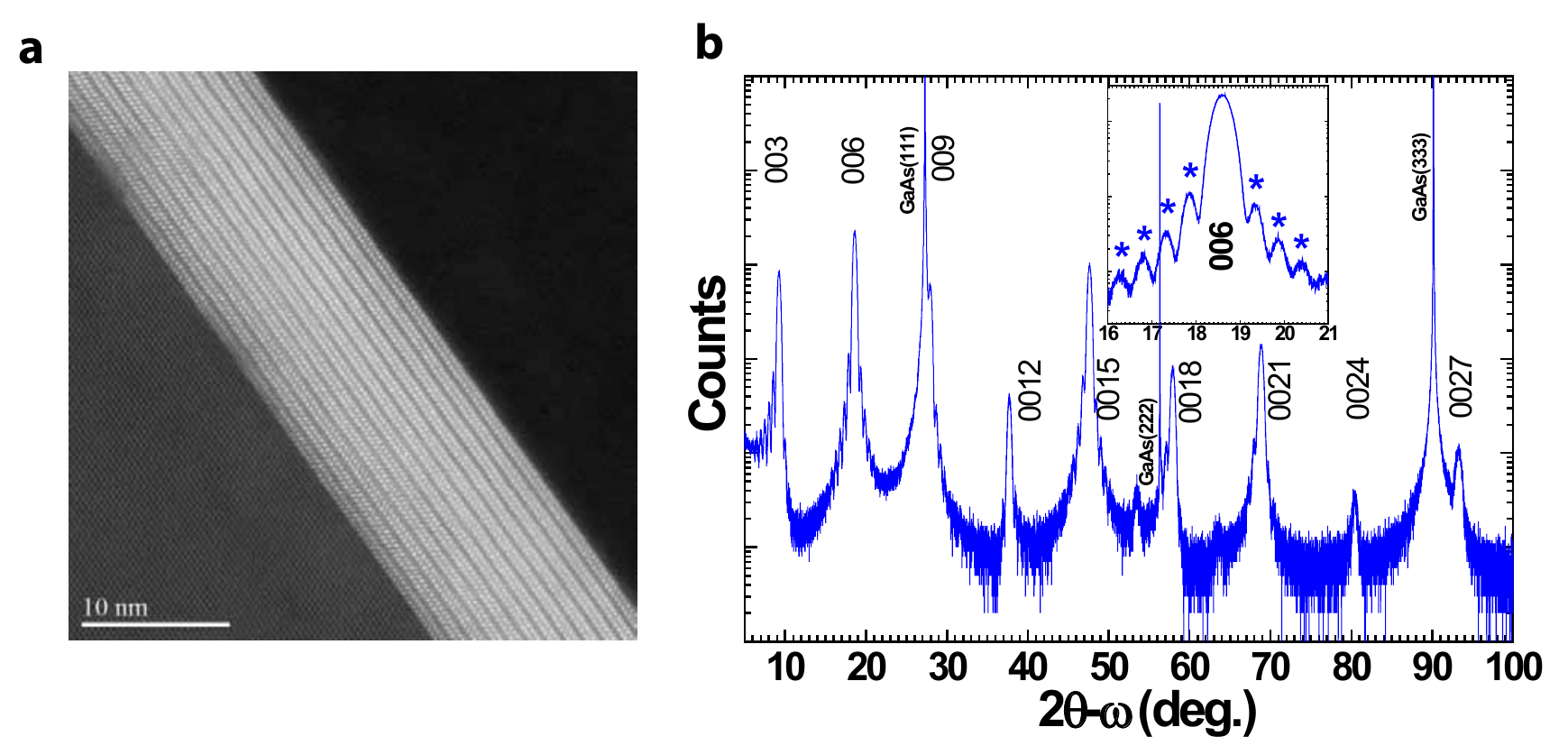}
		\caption{\label{FigS_Structure} \textbf{Structure of the MBE grown Bi$_2$Se$_3$ thin films.} \textbf{a} High-angle annular dark-field scanning transmission electron microscopy (HAADF-STEM) cross-section image of Bi$_2$Se$_3$ film with 12 quintuple layer (~1 QL $\approx 1$~nm). Each QL is delineated by the fringes with darker contrast located at the weak inter-QL bonds in van-der-Waals gap between each QL. \textbf{b} X-ray diffraction spectrum showing the crystalline structure of the films on the GaAs(111) substrate with highly directed {003}-type reflections of the Bi$_2$Se$_3$ film along 111-axis of GaAs. The (006) Bragg reflection is enlarged (in insert) with Kiessig fringes (indicated by stars) which are used to determine the film thickness, about 18 QLs for this film.
		}
	\end{center}
\end{figure*}

\begin{figure*}[h!]
	\begin{center}
		\includegraphics[width=16cm]{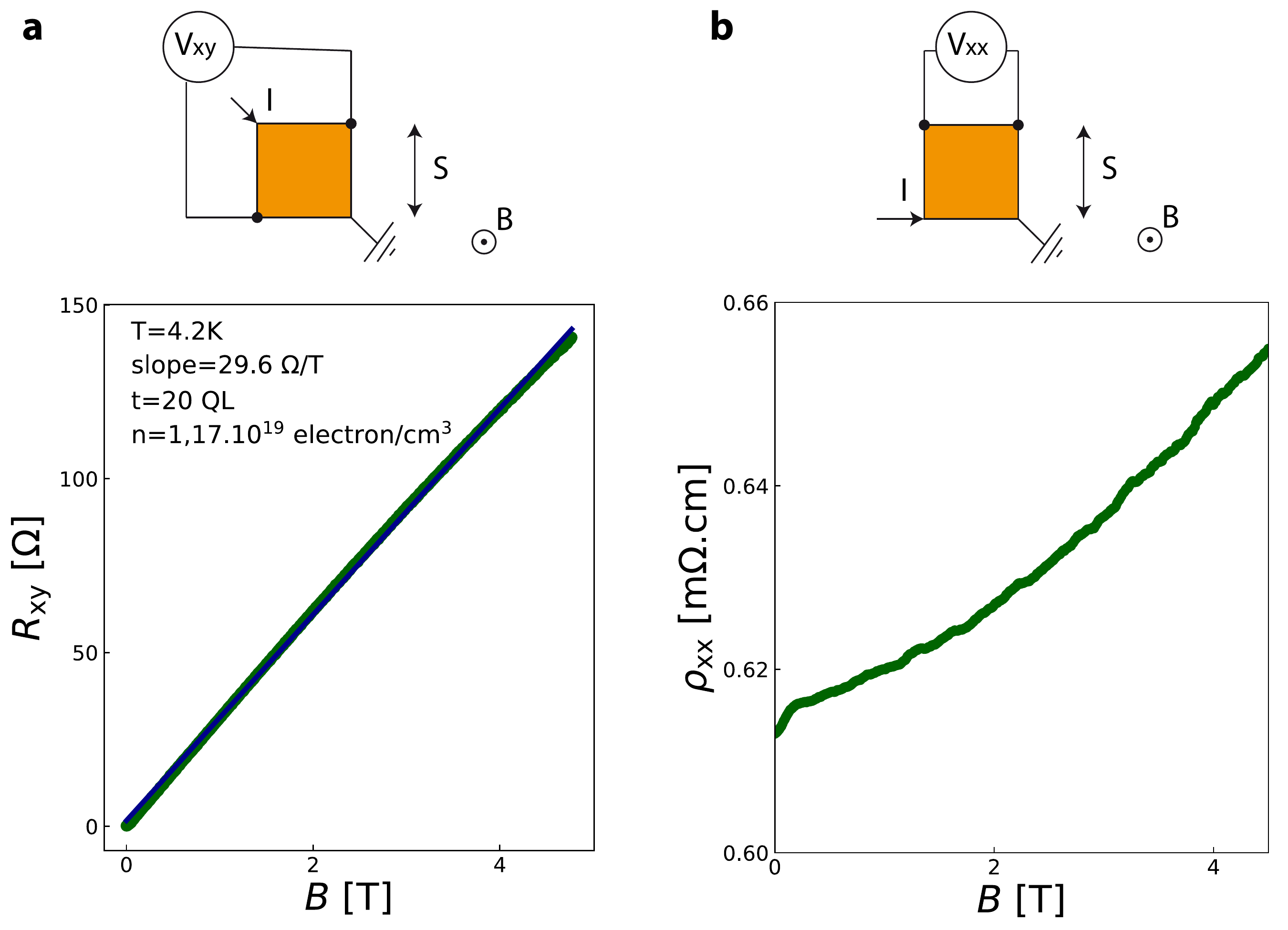}
		\caption{\label{FigS_Transport} \textbf{Transport properties of the Bi$_2$Se$_3$ thin films.} Top : Sketch of electrical connections used for measurements of transport properties in the Van der Pauw geometry. The measurements are realized on microfabricated squares of lateral size 380~$\mu$m. Bottom : The plots show the transverse Hall resistance $R_{\rm H}=29.6~\Omega$.T$^{-1}$ (left panel) and longitudinal resistivity $\rho_{xx}=0.61~$m$\Omega$.cm (right panel) measured at $T=4.2$~K. From the Hall resistance, an electronic density $n=\frac{1}{teR_{\rm H}}=1.2~10^{19}$~electrons.cm$^{-3}$ is determined. Using the effective mass $m^{*}=0.25~m_e$, Ref.~[\cite{Hyde1973-oo}], and the Fermi gas relation between the carrier density and the Fermi velocity, $n=\frac{(m^*v_{\rm F})^3}{3\pi^2\hbar^3}$, one finds $v_{\rm F}=3.2 10^{5}$~m/s. From the Boltzmann relation between the resistivity and the elastic scattering time $\tau$, $\rho_{xx}=m^*/ne^2\tau$, one finds $\tau=0.13~$ps and the elastic mean free path $\ell=v_{\rm F}\tau=42~$nm. Because this distance is smaller than the length of the junction $L=150~$nm, the system is better described in the diffusive regime, where the diffusion constant is obtained from $D=\frac{v_{\rm F\ell}}{3}=40~$cm$^2$s$^{-1}$.
		}
	\end{center}
\end{figure*}

\begin{figure*}[ht!]
	\begin{center}
		\includegraphics[width=16cm]{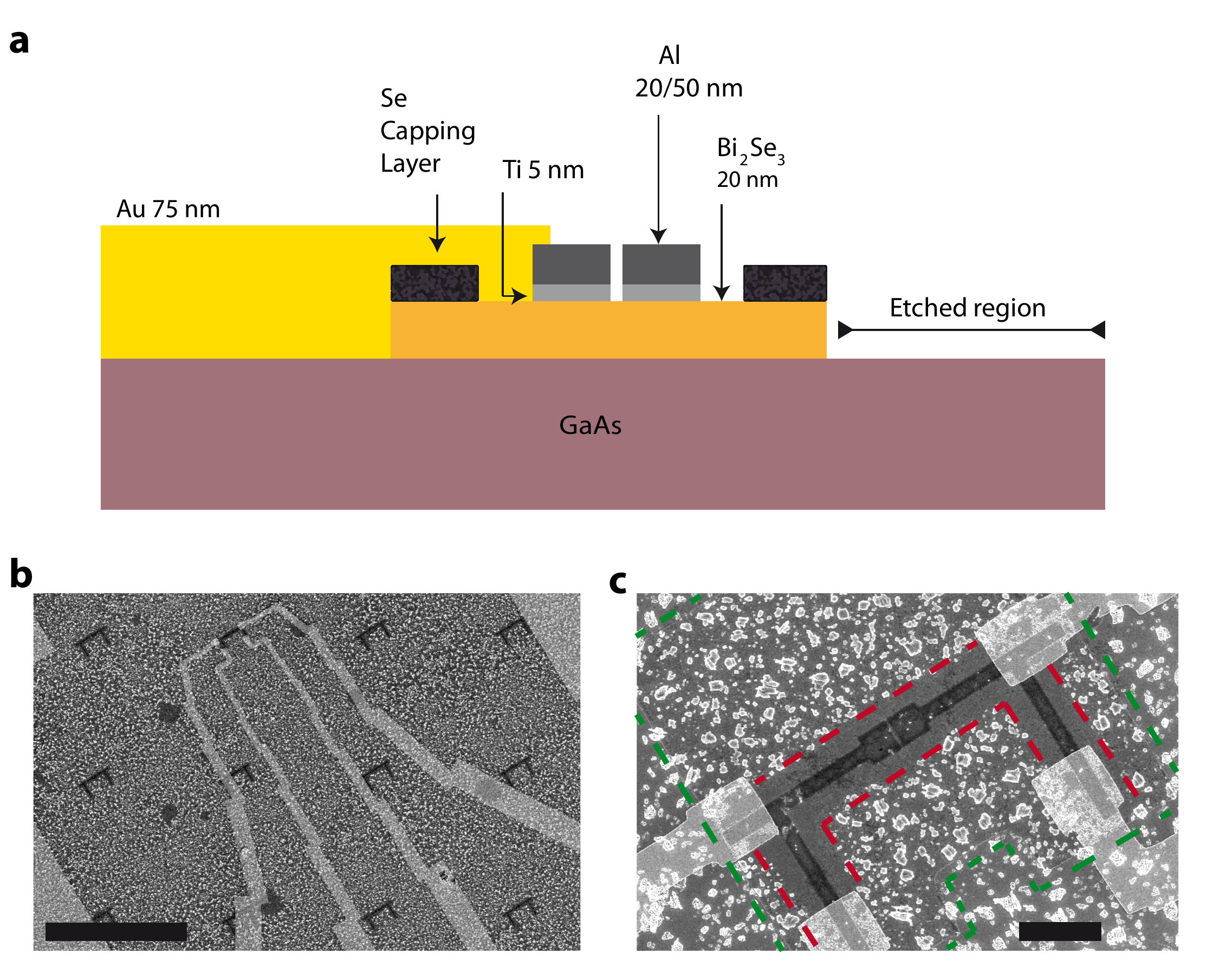}
		\caption{\label{FigS_Devices} \textbf{Sketch and SEM images of devices.} \textbf{a} Lateral view of the device.
		\textbf{b} Scale bar 100~$\mu$m. On these MBE grown films, many junctions are fabricated and examined by SEM lithography. The best junctions are selected and connected by Ti5~nm/Au75~nm electrodes. \textbf{c} Scale bar 10~$\mu$m. The white grains at the surface are Se grains. Before deposition of the Al electrodes (dark area), the Se capping layer is removed by chemical etching in an NMF solution of Na$_{2}$S. Because of the undercut in the PMMA resist, the area on which Se is removed extends beyond the area where the Al electrodes are evaporated. This area where Se is removed is clearly visible and indicated by dashed red lines. After evaporation of the superconducting Al electrodes, the Bi$_2$Se$_3$ film is etched to isolate the junction, the etched contour is visible and highlighted by the dashed green lines.
}
	\end{center}
\end{figure*}
\clearpage

\begin{figure*}[ht!]
	\begin{center}
		\includegraphics[width=16cm]{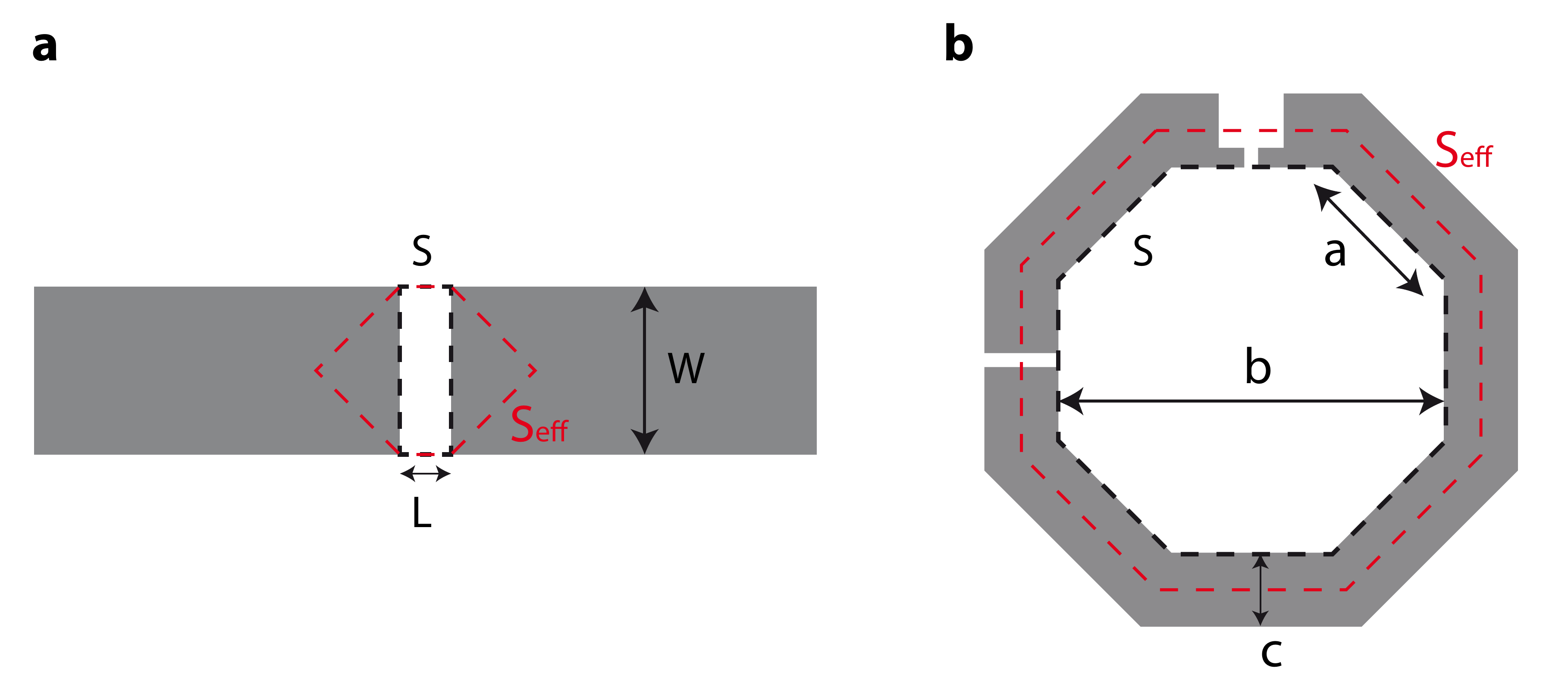}
		\caption{\label{FigS_FluxFocus} \textbf{Effective surface areas due to flux focusing.} \textbf{a} Sketch of a junction of length L and width W. The junction area highlighted by the black dashed lines is $S=LW$. As described in the Supplementary Note 3, the flux lines present within the red dashed lines are diverted into the junction area when the electrodes become superconducting. \textbf{b} Sketch of a the Josephson interferometers used in the main text to detect the anomalous phase shift $\varphi_0$. The junction area is $S=b^2-a^2$, where $a=2.1$~$\mu$m and $b=5$~$\mu$m. Due to the finite width of the electrode, $c=1$~$\mu$m, the effective surface is $S_{eff}=(b+c)^2-(a+c\tan(\pi/8))^2$ and can be re-written $S_{eff}=S+S_{focalized}$ with $S_{focalized}=c(2b+c-2a\tan(\pi/8)-c^2 \tan^2(\pi/8))$.
}
	\end{center}
\end{figure*}

\begin{figure*}[ht!]
	\begin{center}
		\includegraphics[width=16cm]{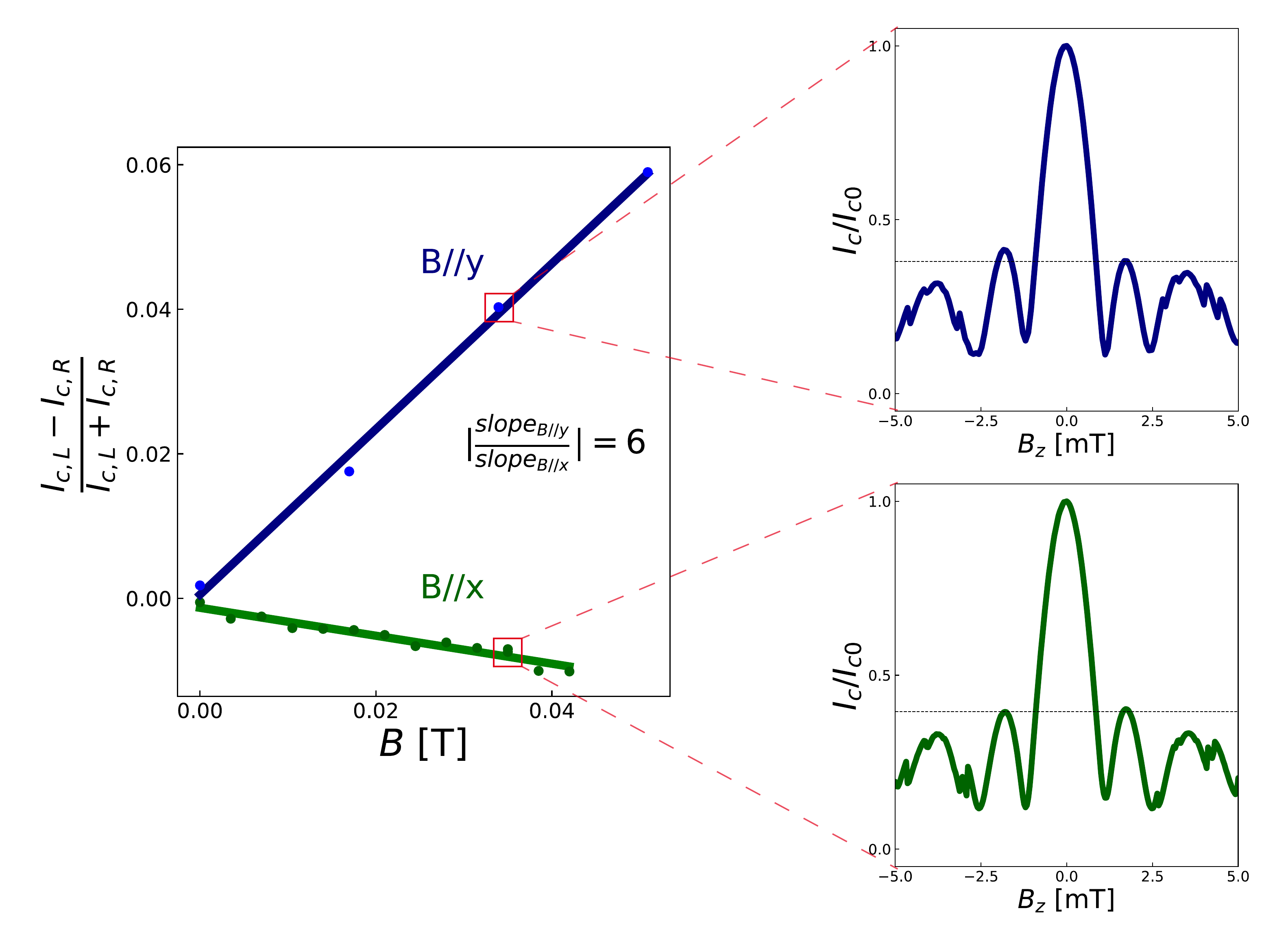}
		\caption{\label{FigS_Direction} \textbf{Fraunhofer pattern as a function of in plane magnetic field orientation.} To quantify the asymmetry, we show the parameter $\frac{I_{\rm c,L}-I_{\rm c,R}}{I_{\rm c,L}+I_{\rm c,R}}$ as a function of the amplitude of magnetic field, where $I_{\rm c,L}$ and $I_{\rm c,R}$ correspond to the critical current of the first lobe at negative and positive $B_z$, respectively. The blue and green curves correspond to in-plane magnetic fields oriented respectively perpendicular and parallel to the current direction. For both orientations, the asymmetry parameter is zero in the absence of in-plane magnetic field. For a finite amplitude of the magnetic field, the asymmetry is much more pronounced when the magnetic field is oriented in the y direction than in the x direction.
}
	\end{center}
\end{figure*}
\clearpage

\begin{figure*}[ht!]
	\begin{center}
		\includegraphics[width=16cm]{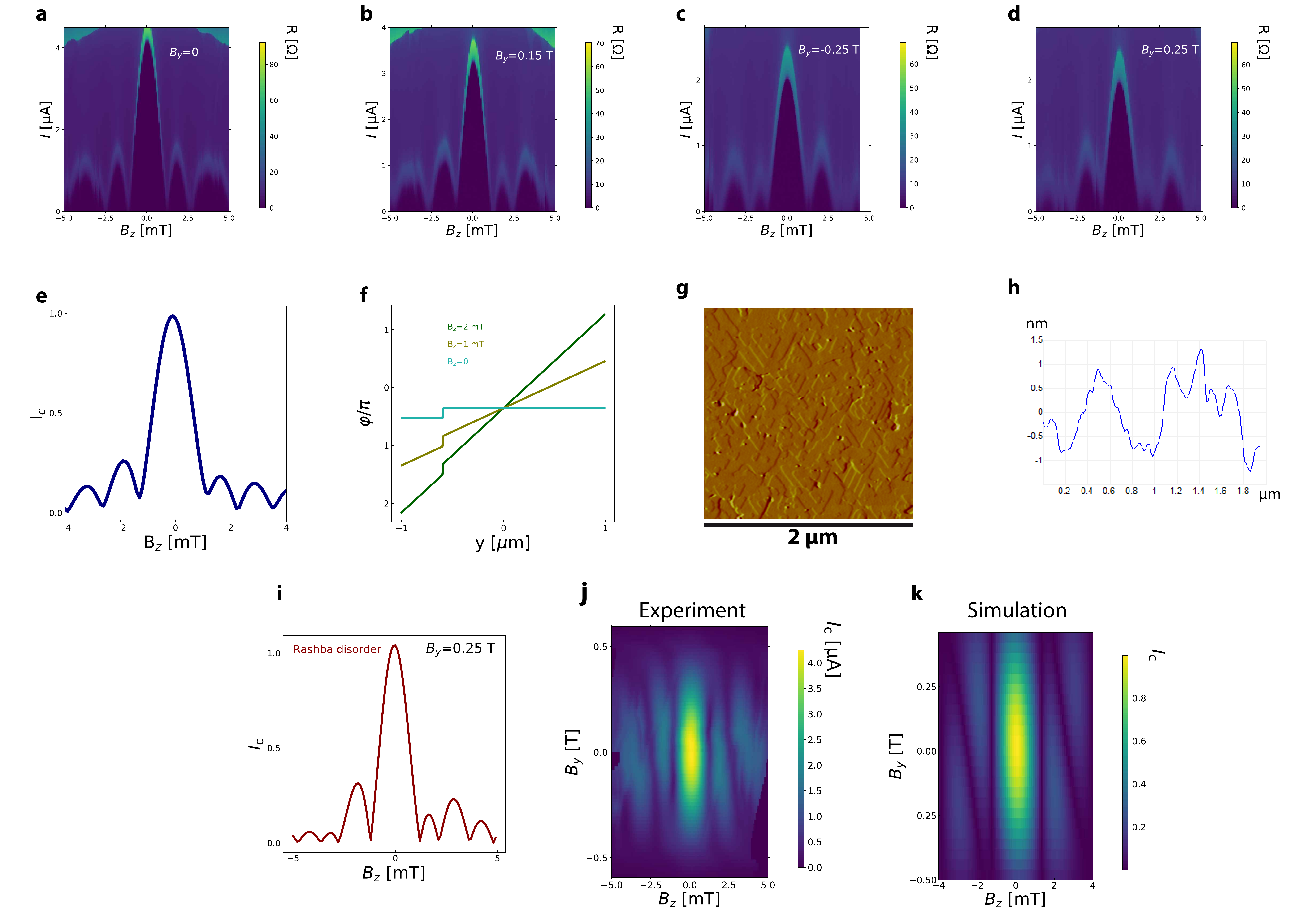}
		\caption{\label{FigS_Simulation} \textbf{Simulation of the anomalous Fraunhofer pattern.} \textbf{abcd} Differential resistance maps $dV/dI$ as a function of current and magnetic field $B_z$ for different values of the in plane magnetic field $B_y$. \textbf{e} An asymmetric Fraunhofer pattern can be generated simply by a jump in the phase difference between the two superconductors along the direction y as plotted panel \textbf{f}. This phase jump can be generated by the in-plane magnetic field in presence of spin-orbit coupling. \textbf{g} 2~$\mu$m $\times$ 2~$\mu$m AFM topographic image of the thin film. \textbf{h} Typical line profile of the sample topography extracted from AFM image. The film thickness changes by $\pm 1$~nm (1 QL) over an average distance of $l_{\rm d}\approx$1~$\mu$m. \textbf{i} Simulation of the Fraunhofer pattern at $B_y$=0.25~T with a Rashba parameter which changes by 0.01~eV\AA\ over a length $l_{\rm d}\approx$1~$\mu$m. \textbf{j} Experimental current map showing the evolution of the asymmetry with the magnetic field. \textbf{k} Simulation of the anomalous pattern using Supplementary Eq.~\ref{eqanpattern}, where phase jumps are induced along the y direction as a consequence of the change in the Rashba coefficient $\alpha$ with the film thickness.}
	\end{center}
\end{figure*}
\clearpage

\begin{figure*}[ht!]
	\begin{center}
		\includegraphics[width=12cm]{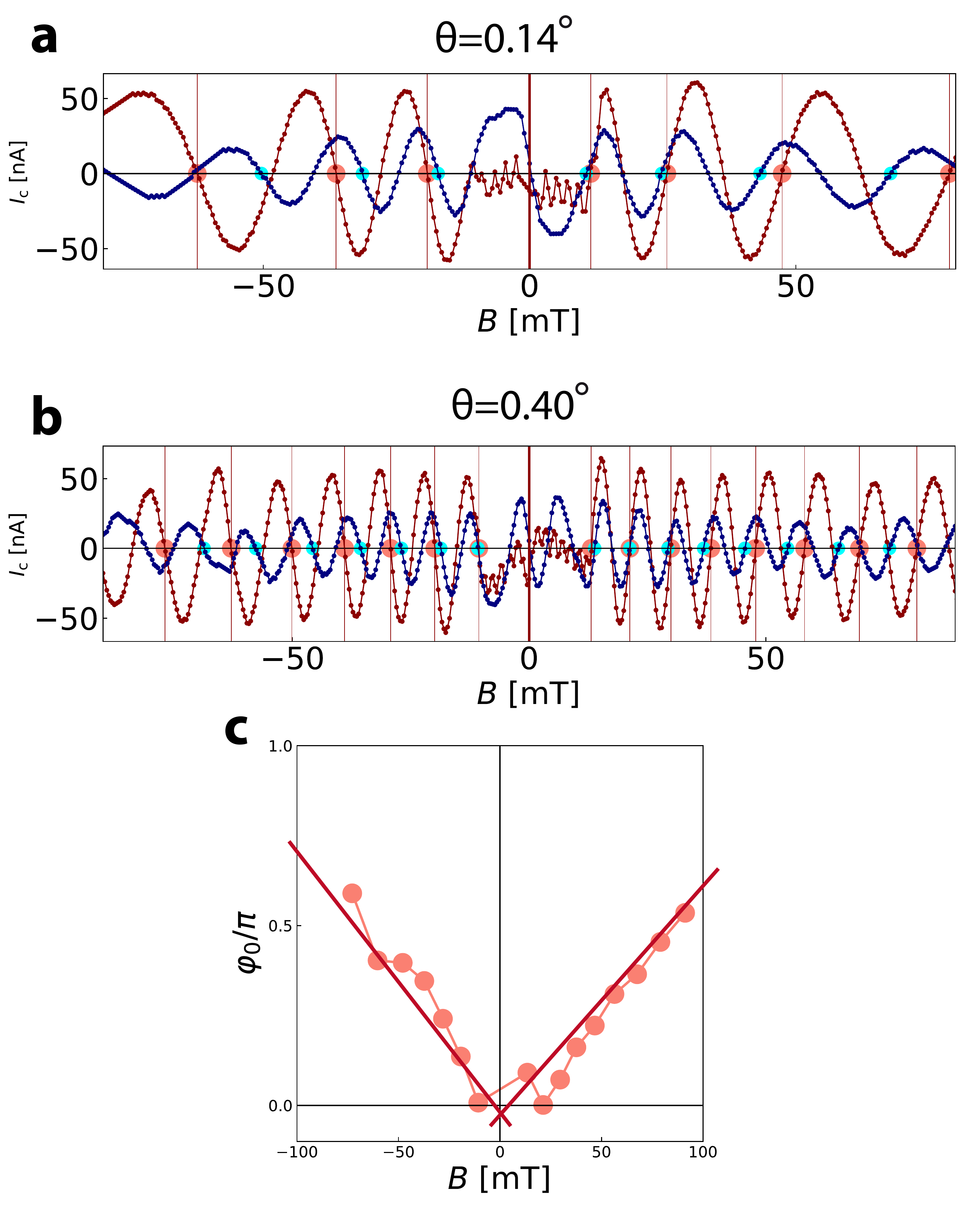}
		\caption{\label{FigS_LowField} \textbf{CPRs near zero magnetic field} \textbf{ab} On these CPRs taken at two different angles of the magnetic field with the sample, the nodes appearing every $2n\pi$ for $n=0,1,2...$ are indicated by red (blue) dots for the reference (anomalous) JI. One can see that the two JIs are in-phase near zero magnetic field, i.e. the first two dots at positive and negative magnetic field, are on the top of each other. As the magnetic field increases, the blue dot shift to lower magnetic field, as a consequence of the anomalous phase shift. From these curves, the phase difference is extracted between each pair of dots(one red, one blue), thus, providing the phase difference between the two JIs as function of main field, shown panel \textbf{c}. This last plot shows that that the two JIs are in-phase at zero magnetic field and reach a dephasing approaching about $\pi/2$ for an in-plane magnetic field of $\simeq 80$~mT.}
	\end{center}
\end{figure*}

\clearpage

\begin{figure*}[ht!]
	\begin{center}
		\includegraphics[width=14cm]{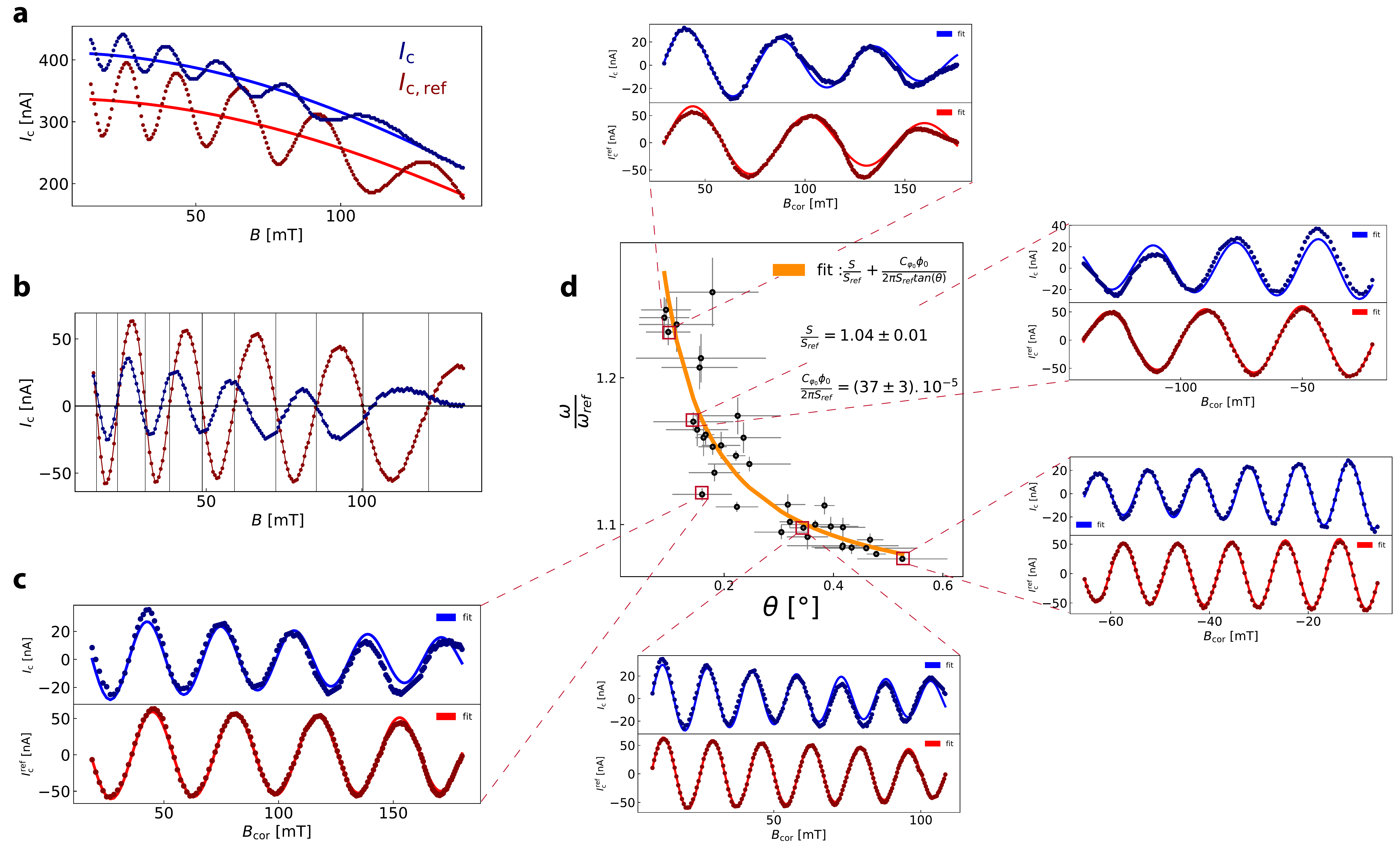}
		\caption{\label{FigS_CPR} \textbf{Comparison of the JIs frequencies as a function of the angle $\theta$: second method.} \textbf{a} The critical current is shown as a function of magnetic field. The red and blue curves correspond to the reference and anomalous device respectively. The anomalous device shows a larger oscillation frequency than the reference device. As the critical current of the large junction decreases with the magnetic field, this leads to a decreasing background fitted for both devices and shown by the continuous lines. \textbf{b} Critical current as function of magnetic field. The background is subtracted to compare precisely the oscillation frequencies of the two devices. The zeros of critical current are not regularly spaced due to flux focusing. \textbf{c} Critical current of the anomalous (upper panel) and reference (lower panel) JI as a function of the corrected magnetic field $B_{\rm cor}$. The zeros of critical current are now regularly spaced. The frequencies for the reference and the anomalous devices are extracted from sinusoidal fit shown as continuous lines. \textbf{d} Ratio of the oscillation frequencies $\frac{\omega}{\omega_{\rm ref}}$ as a function of the angle $\theta$. This ratio diverges as $\frac{1}{\theta}$ for small $\theta$ due to the anomalous phase generated by spin-orbit coupling.}
	\end{center}
\end{figure*}
\clearpage

\section*{Supplementary Tables}

\begin{table*}[ht!]
\begin{center}

\begin{tabular}{|l|c|}
 \hline
 UH$(B_z,\varphi)$U$^\dagger$=H$(-B_z,\varphi)$ & Broken by \\
 \hline
 $\sigma_xP_y$ & $B_y$, $\alpha$, $V_y$  \\
 $\sigma_yP_y$ & $B_x$, $V_y$  \\
 $P_xP_yT$ & $B_x$, $B_y$, $\alpha$, $V_x$, $V_y$  \\
 $\sigma_zP_xP_yT$ & $V_x$, $V_y$  \\

 \hline
\end{tabular}

\caption{\textbf{Symmetry operations U protecting $H(B_z)=H(-B_z)$, from Ref.~[\cite{Rasmussen2016-ct}].} The symmetry operations in the left column are broken by one of the parameters in the right column. These parameters include the in-plane magnetic fields $B_x,B_y$, the asymmetric disorder potentials $V_x,V_y$ and the spin-orbit coefficient $\alpha$. To observe a non symmetric Fraunhofer pattern, all symmetries in the first column need to be broken. One ingredient in the line of the second column is enough to break the corresponding symmetry. For example a non symmetric disorder along y, $V_{y}$, is enough to break all symmetries in the first column and induce a non symmetric Fraunhofer pattern, $I_{\rm c}(B_z)\ne I_{\rm c}(-B_z)$.}
\label{table_paramI}
\end{center}
\end{table*}
\clearpage

\section*{Supplementary Notes}

\subsection{Supplementary Note 1. Details on thin films growth.}

The samples were grown using a multichamber molecular beam epitaxy (MBE) setup at Institut des NanoSciences de Paris. High quality GaAs buffer layers were grown homoepitaxially on GaAs(111) in a Riber compact 21 MBE chamber. The substrates were then transferred under ultrahigh vacuum to a Riber 32 MBE chamber equipped with Bi and Se cells. Growth of  Bi$_2$Se$_3$ epilayers was then carried out at $T=300^{\circ}$C with a Se/Bi beam equivalent pressure ratio higher than 6.5.
The growth temperature and Bi/Se flux ratio were fixed at optimal values within the growth parameters interval leading to high quality epilayers. Such values were determined by monitoring the crystalline quality in-situ by reflection high energy electron diffraction and ex-situ by x-ray diffraction, Supplementary Figure~\ref{FigS_Structure}, and by post growth verification of the electronic structure (Dirac cone fingerprint in angle-resolved photoemission spectra\cite{Vidal2013-tc}). The thickness of the films was checked by x-ray measurements, as shown Supplementary Figure~\ref{FigS_Structure}. Following growth, the samples were capped with a Se protective layer.

\subsection{Supplementary Note 2. A.C. Josephson effect.}

When irradiating a Josephson junction with a microwave signal, its phase oscillates with the frequency $f$ of the applied signal. Standard junctions with a conventional 2$\pi$ periodic CPR, $I_{\rm J}=I_{\rm c}\sin(\varphi)$, will display superfluid current steps, i.e. the Shapiro steps, at the voltage values $V_n=nhf/2e$.

The CPR of topological Majorana states has been predicted to be 4$\pi$ periodic\cite{Snelder2013-cz,Yu_Kitaev2007-ui,Fu2009-uv}, which leads to fractional Shapiro steps. Fractional Shapiro steps have been observed in InSb nanowires driven in the topological regime by a magnetic field \cite{Rokhinson2012-tv}, in the 3D topological insulator HgTe \cite{Wiedenmann2016-xm}, in the 2D topological insulator HgTe quantum well \cite{Bocquillon2017-oc} and in the Dirac semi-metal BiSb \cite{Li2018-mz}.

Bi$_{2}$Se$_{3}$ has been predicted to be a 3D topological insulator due to its strong spin-orbit interaction that leads to an inverted band structure \cite{Zhang2009-ll}. The Dirac cone, which is a predicted characteristic of 3D topological insulators\cite{Zhang2009-ll}, has been observed by photoemission in the thin film used in this study \cite{Vidal2013-tc}.

To investigate the phase periodicity of the supercurrent we illuminate the junction with microwave. This leads to the resistance maps, shown in Fig.~2abcde, as a function of current and microwave power. Regular steps are observed in the IV curve, Fig.~2d, at voltage $V_n=nhf/2e$ where $n$ is an integer and $f$ is the microwave frequency. The behaviour is well captured by the resistively shunted junction (RSJ) model, Fig.~2f upper panel, with a conventional, $2\pi$ periodic, CPR. The absence of 4$\pi$ periodicity is consistent with past experimental works \cite{Kurter2015-fj,Galletti2014-gi,Cho2013-ll,Williams2012-na}. Residual bulk conduction results in a supercurrent mediated predominantly by trivial states which are $2\pi$ periodic in phase.

\textbf{RSJ model}

The A.C. behaviour of a SNS Josephson junction can be captured by the RSJ model. The equivalent circuit of a SNS junction consists of a Josephson junction of critical current $I_{\rm c}$ in parallel with a resistor with a resistance $R_{\rm N}$. The microwave signal is added in the model in the form of an ac current $I_{\rm RF}\sin(2\pi ft)$, with $f$ the microwave frequency.
The current in such a circuit is:

\begin{equation}
I=\frac{h}{2eR_{\rm N}}\frac{d\varphi}{dt}+I_{\rm c}\sin(\varphi)+I_{\rm RF}\sin(2\pi ft)
\label{eqRSJ}
\end{equation}

This equation is solved numerically for each pair of $I$ and $I_{\rm RF}$. The voltage V is found from the second Josephson relation by a time average $\langle \rangle_t$ of the phase solution derivative:

\begin{equation}
V=\frac{h}{2e}\langle\frac{d\varphi}{dt}\rangle_t
\end{equation}

A 4$\pi$ periodic contribution to the supercurrent can be added by using a CPR of the form $I_{\rm J}=I_{\rm c}(\frac{4}{5}\sin(\varphi)+\frac{1}{5}\sin(\varphi/2))$. The result is shown in the lower panel of Fig.~2f. 

The RSJ model has been extended to understand the role of thermal effect which can mask the $4\pi$ periodic contribution to the supercurrent in Bi$_2$Se$_3$\cite{Le_Calvez2018-ku}.


\clearpage

\subsection{Supplementary Note 3. Periodicity of the Fraunhofer and Josephson interferometers.}

The first node of the Fraunhofer patterns measured on the junction described in the main text, Fig.~3, and the junctions described in Supplementary Figure~\ref{FigS_Direction} and Supplementary Figure~\ref{FigS_Simulation}abcd, are located at the value $B_0\simeq1.2$~mT. These two junctions have identical widths $W$=2~$\mu$m and electrode spacing $L=150$~nm. The field position of the first node is expected at $B_0=\frac{\phi_0}{W(L+2\lambda_{z})}$.

For the aluminum electrodes, the effective penetration depth $\lambda_{\rm eff}$ is given by:
\begin{equation}
\lambda_{\rm eff}=\lambda_{\rm L}\sqrt{1+\frac{\xi}{\ell}}
\end{equation}
where $\lambda_{\rm L}=16~$nm is the bulk London penetration depth of aluminum,   $\ell$ is the mean free path and $\xi$ is the coherence length\cite{Larkin_undated-gy} calculated from:


\begin{equation}
\xi=0.36\sqrt{\frac{3\hbar v_{\rm F} \ell}{2k_{\rm B} T_{\rm c }}}
\end{equation}

From the measured critical temperature $T_{\rm c}=0.4$~K of the aluminum titanium bi-layer and mean free path of aluminum $\ell=\frac{\sigma m v_{\rm F}}{ne^2}=50$~nm, we find $\xi=610~$nm and $\lambda_{\rm eff}=58$~nm.

For a magnetic field $B_z$ perpendicular to a thin film, the penetration depth is given by\cite{Gubin2005-wf}:

\begin{equation}
\lambda_{z}=\lambda_{\rm eff} \coth(\frac{d}{\lambda_{\rm eff}})
\end{equation}

For a film of thickness $d=20$~nm, we find $\lambda_{z}=175~$nm. With this value of London penetration depth, we find $B_0\simeq 2$~mT, which is about 2 times larger than the experimental value.  

This discrepancy can be explained by flux-focusing which increases the magnetic field in the junction. A simple way to take this effect into account is described in Ref.\cite{Molenaar2014-vg}. When the electrode becomes superconducting, part of the magnetic flux lines that would penetrate the electrodes in the normal state are now expelled from the electrodes and focalized into the junction area. We can estimate the amount of flux focusing by considering the shortest distance a flux line has to be diverted to not pass through the superconducting electrode. This leads to an increase of the effective junction area from $S$ to $S_{eff}=S+S_{focalized}$ where $S_{focalized}=2\times(W/2-\lambda_z)^2$ as sketched Supplementary Figure~\ref{FigS_FluxFocus}. Taking into account this flux-focusing yields $B_0\approx$~0.9~mT, which is close to the experimental value.


In Supplementary Figure~\ref{FigS_Direction} and Supplementary Figure~\ref{FigS_Simulation}, abrupt jumps of the critical current magnitude are observed for $B_z\gtrsim 3$~mT. Such jumps are expected in presence of trapped vortices. Similar behavior has been observed in Josephson junctions fabricated with type II superconducting electrodes\cite{Kim2017-cc}. In our case, despite the modified London penetration depth and coherence length in our thin films, we have $\xi \approx 10\lambda_{\rm eff}$, which implies that the aluminum remains of type I and no vortices are expected in the electrodes.

However, in hybrid Josephson junctions, the effective penetration depth in the semiconducting material is expected to be much larger than for the superconducting electrodes because the carrier density in the semiconductor is much smaller than in the electrodes. Thus, it is quite plausible that the hybrid heterostructure aluminum/semiconductor becomes a type II superconductor. This probably explains the ubiquitous observation of flux-trapping effects in hybrids SNS junctions made from semiconducting materials\cite{Kim2017-cc,Williams2012-na}.

As seen in Fig.~4, the period of the Josephson interferometers increase with in-plane magnetic field. The major ingredients for explaining the increasing period with magnetic field are first, the existence of flux-focusing and second, the increasing penetration depth with in-plane magnetic field.

As described above, flux-focusing leads to an increase of the effective junction area and so to a correspondingly short period. However, the increase in the penetration depth leads to a reduction in flux focusing as the magnetic lines that were diverted into the junction area are now penetrating the electrodes. This leads to a decrease of the effective area of the junction and so to an increase of the period. For example, in the Fraunhoffer pattern shown Fig.~\ref{FigS_Direction}, the field position of the second critical current node appears at a larger magnetic field than expected from the field position of the first node. This effect has also been seen in other recent works\cite{Williams2012-na,Suominen2017-ls}. This phenomena also occurs in the Josephson interferometers. As sketched Supplementary Figure~\ref{FigS_FluxFocus}, because of the finite width of the superconducting electrode making the interferometer, the effective area of the interferometer is larger than the inner area.  Upon increasing the magnetic field, the increase in the penetration depth leads to a reduction of the effective area and so to an increase of the oscillation period of the interferometer, as shown Fig.~4, Fig.~5, Supplementary Figure~\ref{FigS_LowField} and Supplementary Figure~\ref{FigS_CPR}.

The increase of the penetration depth with magnetic field could be due to the penetration of vortices in the electrodes or to a reduction in the amplitude of the superconducting order parameter at the approach of the upper critical field. According to Ref.~\cite{Tinkham1996-zb}, the penetration depth increases as $\lambda_eff\propto1/\Delta(H)\rightarrow\infty$ for an in-plane magnetic field.

Finally, one also want to estimate the contribution from the circulating superfluid current, of maximum amplitude given by the critical current $I_c\simeq1~\mu$A, to the magnetic flux in the junction or JI. 

For a Josephson junction, the screening of the applied magnetic field generated by the Josephson supercurrent is negligible when the characteristic size of the junction is smaller than the Josephson length:

\begin{equation}
\lambda_J=\left(\frac{\hbar}{2e\mu_0(L+2\lambda_z)J_0}\right)^{1/2} 
\end{equation}

where $J_0$ is the current density.
From this relation, one finds $\lambda_J\simeq5~\mu$m, which is larger than the characteristic lengths of the Josephson junctions where the large junctions have a size $2~\mu$m$\times150~$nm and the small junctions have a size $150$~nm$\times150~$nm. This implies negligible contribution of the Josephson supercurrent to the magnetic flux in the junctions.

For a JI, the actual maximum flux is given by $\phi=\phi_{applied}-L.I$, where $L$ is the inductance of the JI loop. We estimate that the inductance of the JI loop is about $L\simeq 1.10^{-11}$~H. From this inductance, one finds $LI_c/\phi_0=10^{-2}$, showing that the superconducting current contribution to the total magnetic flux is negligible in comparison to the flux due to the applied magnetic field.

\subsection{Supplementary Note 4. Simulation of the asymmetric Fraunhofer pattern.}

A general relation for the critical current as a function of perpendicular $B_z$ and in-plane magnetic field $B_y$ is given by:

\begin{equation}
   I_{\rm c}(B_z,B_y)=\max_{\varphi'}\int_{-W/2}^{W/2}\int_{0}^{d} j_0\sin(\varphi'+\varphi_{xy}(y)+\varphi_{xz}(z)+\varphi_0(y))\mathrm{d}y \mathrm{d}z
\label{eqanpattern}
\end{equation}

where $j_0$ the critical current density, $d\simeq 20$~nm the thickness of the film, $\varphi'$ an arbitrary global phase shift, $\varphi_{xy}(y)=\frac{2\pi (L+2\lambda_{z})yB_z}{\phi_0}$ and $\varphi_{xz}(z)=\frac{2\pi (L+2\lambda_{\rm eff})zB_y}{\phi_0}$, are the magnetic fluxes produced in the junction by the magnetic field $B_z$ and $B_y$ respectively, where $L=150$~nm is the distance between the superconducting electrodes.

Finally, the last phase argument $\varphi_0(y)$ in Supplementary Eq.~\ref{eqanpattern} is the anomalous phase which is allowed to depend on y due to disorder\cite{Bergeret2015-so}:
\begin{equation}
    \varphi_0=\frac{\tau m^{*2} E_{\rm Z}(\alpha L)^3}{3\hbar^6 D}
\label{Eq:phi0}
\end{equation}

where $\tau=0.13~$ps is the elastic scattering time, $D=40~$cm$^2$s$^{-1}$ is the diffusion constant, $m^*=0.25~m_e$ is the effective electron mass, $\alpha(y)$ is the spin-orbit coefficient that depends on disorder, $E_{\rm z}=\frac{1}{2}g\mu_{\rm B}B$ is the Zeeman energy with $g=19.5$\cite{Wolos2016-lj}. As shown by photoemission, the Rashba parameter depends on the thickness of the film\cite{Zhang2010-kf}. Using Supplementary Eq.~\ref{Eq:phi0}, one finds that a change of the Rashba coefficient by $\Delta\alpha=0.01$~eV.\AA ~leads to a phase jump of $\Delta\varphi_0\simeq~0.3 \pi$. This phase jump along the y-direction is sketched Supplementary Figure~\ref{FigS_Simulation}f and is sufficient to generate an asymmetric Fraunhofer pattern, as shown in Supplementary Figure~\ref{FigS_Simulation}e. 

The mere existence of a large value for $\alpha=0.38$~eV\AA~does not lead to an asymmetric Fraunhofer pattern, in the absence of disorder, as the induced anomalous phase-shift $\varphi_0$ can always be compensated by the arbitrary phase $\varphi'$. In other words, because in Supplementary Eq.\ref{eqanpattern} the critical current is obtained by maximizing over the arbitrary phase $\varphi'$, a global change of $\varphi_0$ will be compensated by an equivalent change of the arbitrary phase $\varphi'$. Only anomalous phase jumps along the y direction can generate an asymmetric Fraunhofer pattern, in agreement with Supplementary Table 1, indicating that finite disorder $V_y$ must be present for the asymmetry to be present.

AFM topographic images of our MBE films, Supplementary Figure~\ref{FigS_Simulation}gh, show that the film thickness changes by an amount $\pm$1~nm over a length $l_{\rm d}=1$~$\mu$m. We model the variation in the spin-orbit coefficient by the function $\alpha(y)=\alpha_0+\Delta \alpha \sin(2\pi y/l_{\rm d})$ where $\Delta \alpha=0.01$~eV\AA. The resulting Fraunhofer pattern is shown in Supplementary Figure~\ref{FigS_Simulation}hk, displaying good agreement with the experimental data.

\subsection{Supplementary Note 5. Comparison of the JIs frequencies as a function of the angle $\theta$ (second method).}

In addition to the method described in the main text for extracting the frequency ratio $\frac{\omega}{\omega_{\rm ref}}$ between the two interferometers, we present here a second method based on a rescaling of the applied magnetic field. To compare the frequencies of the two interferometers we first remove the critical current background. This is done by fitting the $I_{\rm c}(B)$ curve of the two devices with a parabola, Supplementary Figure~\ref{FigS_CPR}a, which is removed from the experimental data and shown Supplementary Figure~\ref{FigS_CPR}b. 

The oscillation frequency of the reference interferometer should not vary with the amplitude of the magnetic field applied with a small angle $\theta$ with the plane of the sample. However, because of flux focusing effects, this frequency is observed to change with the the magnetic field, Supplementary Figure~\ref{FigS_CPR}b.

To correct for this flux focusing effect, we rescale the magnetic field, giving the corrected magnetic field scale B$_{\rm cor}$, such that the reference signal becomes periodic as shown in Supplementary Figure~\ref{FigS_CPR}c. This same corrected scale is applied to the anomalous device. In this corrected field scale, the frequency ratio $\frac{\omega}{\omega_{\rm ref}}$ is extracted and plotted Supplementary Figure~\ref{FigS_CPR}d. At large $\theta$, this ratio is equal to the ratio of areas $S/S_{\rm ref}\simeq 1$, however, for small $\theta$, this ratio increases as $1/\tan(\theta)$, indicating the presence of an anomalous phase shift $\varphi_0=C_{\varphi_0}B$.

In Supplementary Figure~\ref{FigS_CPR}b and in the main text, the angle $\theta$ is determined from the last oscillation period $\Delta$B of the reference device by the relation $\theta=\arcsin(\frac{\phi_0}{S\Delta B})$. The error on the determination of the angle $\delta\theta$ is due to the error $\delta(\Delta B)$ which is estimated as the difference between the last oscillation period and the before last oscillation period. The error on the angle is given by the relation $\delta \theta=\frac{\phi_0\delta(\Delta B)}{S\Delta B^{2}\sqrt{1-\left(\frac{\phi_0}{S\Delta B}\right)^2}}$.

\section*{Supplementary References}

\bibliographystyle{naturemag}
\bibliography{Bibliography}